\documentclass[11pt, letterpaper]{article}

\usepackage[left=1in, right=1in, top=1in, bottom=1in]{geometry}
\usepackage[utf8]{inputenc}
\usepackage[T1]{fontenc}
\usepackage{bm}
\usepackage{type1cm}
\usepackage{lettrine}
\usepackage{amsmath,amssymb,amsthm}
\usepackage{moreverb}
\usepackage{mathtools}
\usepackage{amsmath}
\usepackage{svg}

\usepackage{amssymb}
\usepackage{algorithmic}
\usepackage{graphics}
\usepackage{graphicx}
\usepackage{subfigure}
\usepackage{caption}
\usepackage{extarrows}
\usepackage{color}
\usepackage{framed}
\usepackage{wrapfig}
\usepackage{algorithm}
\usepackage{algorithmic}
\usepackage{bm}
\usepackage{mathrsfs}
\usepackage{mathabx}
\usepackage{multirow}
\usepackage{longtable}
\usepackage{hyperref}
\usepackage{paralist}
\usepackage{indentfirst}
\usepackage{relsize}
\usepackage{extarrows}
\usepackage{upgreek}
\usepackage{bm}
\usepackage{mwe}
\usepackage[dvipsnames,table]{xcolor}
\usepackage{booktabs}
\usepackage{authblk}
\usepackage{lettrine}
\usepackage{type1cm}
\usepackage{threeparttable}
\usepackage[sort&compress,numbers]{natbib}
\usepackage[figurename=Figure]{caption}
\usepackage[normalem]{ulem}
\usepackage{adjustbox}
\usepackage{ulem}
\usepackage{xcolor}

\usepackage{amsmath,amsfonts,bm}

\newcommand{\rev}[1]{\textcolor{black}{#1}}

\def\eqref#1{equation~\ref{#1}}

\def\1{\bm{1}}

\DeclareMathAlphabet{\mathsfit}{\encodingdefault}{\sfdefault}{m}{sl}
\SetMathAlphabet{\mathsfit}{bold}{\encodingdefault}{\sfdefault}{bx}{n}

\graphicspath{ {./Figure/} }
\usepackage[font=footnotesize,labelfont=bf]{caption}

\newcommand{\graph}[1]{\mathcal{}}
\providecommand{\keywords}[1]{\textbf{\textit{Keywords: }} #1}

\hypersetup{
bookmarks=true,
bookmarksopen=true,
bookmarksnumbered=true,
unicode=false,
pdftoolbar=true,
pdfmenubar=true,
pdffitwindow=false,
pdfstartview={FitH},
pdftitle={My title},
pdfauthor={Author},
pdfsubject={Subject},
pdfcreator={Creator},
pdfproducer={Producer},
pdfkeywords={keywords},
pdfnewwindow=true,
colorlinks=true,
linkcolor=blue,
citecolor=blue,
filecolor=blue,
urlcolor=blue
}

\begin{document}
\title{\textbf{PhononBench:A Large-Scale Phonon-Based Benchmark for Dynamical Stability in Crystal Generation}}


\author[1]{Xiao-Qi Han\textcolor{black}{$^\dagger$}}
\author[1,*]{Ze-Feng Gao\textcolor{black}{$^\dagger$}}
\author[1]{\textcolor{black}{Wen-Kao Li}}
\author[1]{Peng-Jie Guo}
\author[1,*]{Zhong-Yi Lu}

\affil[1]{\small School of Physics, Renmin University of China, Beijing, China}

\affil[*]{Corresponding authors\vspace{12pt}}

\date{}

\maketitle

\normalsize

\vspace{-28pt} 
\begin{abstract}
\small

In recent years, generative artificial intelligence has made significant advances in the design of crystalline materials, giving rise to a variety of approaches based on graph neural networks, diffusion models, and large language models. Existing evaluations commonly follow the stability–uniqueness–novelty (S.U.N.) framework, where ``stability'' is primarily assessed using thermodynamic criteria, which do not fully capture the dynamical stability essential for a material’s practical existence. In fact, dynamical stability is a key determinant of whether a material can be synthesized and persists, with phonon spectrum calculations based on first-principles methods serving as the established standard for its evaluation. However, the high computational cost of such calculations has, until now, prevented large‑scale and systematic assessment of dynamical stability in generated crystals. In this work, we introduce PhononBench, the first large-scale benchmark for dynamical stability in AI-generated crystals. Leveraging the recently developed MatterSim interatomic potential, which achieves density-functional-theory (DFT)–level accuracy in phonon predictions across more than 10,000 materials, PhononBench enables efficient large-scale phonon calculations and dynamical-stability analysis for \textcolor{black}{133,838} crystal structures generated by \textcolor{black}{7} leading crystal generation models. PhononBench reveals a widespread limitation of current generative models in ensuring dynamical stability: \textcolor{black}{unless otherwise specified, all reported dynamical-stability metrics are evaluated at a phonon-frequency threshold of $-0.1$~THz}, with the average dynamical-stability rate across all generated structures being only \textcolor{black}{32.15\%}, and the top-performing model, MatterGen, reaching just \textcolor{black}{45.05\%}. Further case studies show that in property‑targeted generation—illustrated here by band‑gap conditioning with MatterGen—the dynamical‑stability rate remains as low as \textcolor{black}{41.0\%} even at the optimal band‑gap condition of 0.5 eV. In space‑group‑controlled generation, higher‑symmetry crystals exhibit better stability (e.g., cubic systems achieve rates up to \textcolor{black}{53.4\%}), yet the average stability across all controlled generations is still only \textcolor{black}{44.7\%}. An important additional outcome of this study is the identification of \textcolor{black}{32,995} crystal structures that are phonon-stable across the entire Brillouin zone \textcolor{black}{under the strict threshold of $-0.001$~THz}, providing a substantial pool of reliable candidates for future materials exploration. By establishing the first large‑scale dynamical‑stability benchmark, this work systematically highlights the current limitations of crystal generation models and offers essential evaluation criteria and guidance for their future development toward the design and discovery of physically viable materials. All model-generated crystal structures, phonon calculation results, and the high-throughput evaluation workflows developed in PhononBench have been openly released at \url{https://github.com/xqh19970407/PhononBench}. \textcolor{black}{In addition, a web-based service is accessible at \url{http://phononbench.cn/}, enabling minute-level ultra-fast phonon predictions.}
\end{abstract}

\vspace{12pt} 

{\footnotesize \textcolor{black}{$^\dagger$These authors contributed equally to this work.}}
\section*{Introduction}\label{sec1}


In recent years, AI-driven inverse materials design has advanced rapidly and comprehensively, achieving breakthroughs across multiple aspects of crystal materials design~\cite{mattergen,gnome,InvDesFlow,invdesreview}. Diffusion-based approaches~\cite{ddpm,song2021Score-based}, such as DiffCSP~\cite{diffcsp}, have demonstrated significant improvements in both speed and accuracy for crystal structure prediction compared with traditional DFT-based search workflows. CrystalFlow~\cite{CrystalFlow} further enhances efficiency by reformulating the underlying algorithm using flow matching, substantially accelerating inference. For functional materials design, MatterGen~\cite{mattergen} exhibits strong performance across key metrics, including Stability, Uniqueness, Novelty~(short as S.U.N.), and leverages adapter-based fine-tuning to enable precise generation of various classes of functional materials. The active learning–based workflow for inverse functional materials design (InvDesFlow-AL)~\cite{InvDesFlow-AL} has achieved notable success in the design of high-temperature superconductors, together with other generative approaches~\cite{cdvae,condvae,cond-cdvae} for functional materials, further demonstrating the capability of AI to explore complex material systems. Meanwhile, crystal generation models incorporating space-group control—such as CrystalFormer~\cite{crystalformer,crystalformer-RL} and DiffCSP++~\cite{diffcsp-pp} significantly improve the symmetry and physical plausibility of generated structures. With the rapid development of large language models~\cite{openai2024gpt4technicalreport,DeepSeek-R1}(LLM), methods such as CrystaLLM~\cite{crystallm} and FlowLLM~\cite{flowllm} integrate natural language into crystal generation, enabling more intuitive and flexible conditioning for materials design.

However, it is essential to recognize the limitations of current progress. Despite the rapid development of crystal generative models, their assessment of material stability has largely focused on thermodynamic stability, typically evaluated using metrics such as $E_{\text{hull}}$ (energy above the convex hull)~\cite{matbench,InvDesFlow-AL,mattergen,dpa2}. Yet the synthesizability and practical existence of materials depend not only on thermodynamic stability but, more critically, on dynamical stability—whether a structure resides in a local potential well and can withstand small perturbations without collapsing. Dynamical instability often manifests as imaginary phonon modes~\cite{instability,InvDesFlow}, which not only directly signal mechanical instability but have also long posed a persistent challenge for DFT practitioners~\cite{RevModPhys-Phonons1}. The origins of imaginary modes are diverse, potentially arising from symmetry breaking, insufficient q-mesh resolution, pseudopotential choices, approximation errors, or intrinsic structural instability~\cite{TOGO20151,RevModPhys-Phonons2}. More importantly, because rigorous dynamical stability validation (e.g., full phonon dispersion calculations) is computationally demanding~\cite{PhysRevX-Phonons3}, most existing generative models have never performed systematic dynamical stability tests on their generated structures. As a result, these models may produce a large number of structures that appear thermodynamically stable but are in fact dynamically unstable~\cite{Peng2025}, undermining both the reliability and practical utility of their predictions. Therefore, building on the successes of current generative models, establishing systematic and efficient dynamical stability evaluation for generated structures has become a key challenge and an important frontier for guiding computationally generated materials toward experimental synthesis and enhancing the trustworthiness of model predictions.

With the rapid development of universal machine-learning interatomic potentials (uMLIPs)~\cite{mattersim,dpa2,SevenNet,Deng2023CHGNet,nequip}, a growing number of models have achieved energy and force prediction accuracies that approach or even surpass those of DFT, thereby greatly enhancing computational efficiency in atomistic materials simulations~\cite{equiformerv2,escn,Orb,Orb-v3}. Representative advances include MEGNet~\cite{MEGNet}, which reaches a formation energy accuracy of 21 meV on the GNoME~\cite{gnome}, and M3GNet~\cite{m3gnet}, which incorporates atomic coordinates, lattice vectors, and three-body interactions and has become one of the most prominent uMLIPs. Building on M3GNet, MatterSim~\cite{mattersim} is pretrained on 17 million first-principles data points, enabling zero-shot generalization across the first 89 elements of the periodic table over temperatures of 0–5000 K and pressures of 0–1000 GPa. Its accuracy in predicting energies, forces, and stresses surpasses previous models such as MACE~\cite{Batatia2022mace} by roughly an order of magnitude, enabling reliable calculations of lattice dynamical, mechanical, and thermodynamic properties. More importantly, Miguel A. L. Marques et al. conducted a systematic assessment based on phonon calculations for over 10,000 materials (Fig.~\ref{Benchmark-main})~\cite{PhonnBench}. Their results show that MatterSim achieves phonon-spectrum prediction accuracy comparable to DFT, with average errors even smaller than the inherent differences between the PBE and PBEsol functionals, and markedly outperforming all other models. Furthermore, in dynamical-stability classification, MatterSim attains a 95\% true-positive rate, achieving a level of reliability nearly equivalent to a full DFT workflow while requiring only a tiny fraction of its computational cost. Taken together, these results clearly demonstrate that MatterSim is not merely “good enough,” but in fact a genuinely reliable and efficient tool for high-throughput phonon calculations and dynamical-stability analysis.

In this work, we introduce PhononBench and systematically perform phonon calculations for  \textcolor{black}{133,838} crystal structures generated by \textcolor{black}{7} commonly used generative models, providing the first large-scale assessment of the dynamical stability of AI-generated materials. PhononBench results reveal that current generative models remain markedly limited in ensuring phonon stability: the average dynamical stability rate across all models is only \textcolor{black}{32.15\%}, with the best-performing MatterGen model reaching merely \textcolor{black}{45.05\%}. Further case studies on functional-material generation, exemplified by band-gap–conditioned generation with MatterGen, show that although the highest stability is achieved at a target band gap of 0.5 eV, the corresponding dynamical stability rate remains low at \textcolor{black}{41.0\%}. In space-group–controlled generation, crystals with higher symmetry exhibit improved phonon stability—for instance, cubic systems reach stability rates of up to \textcolor{black}{53.4\%}—yet the overall average stability is still modest at \textcolor{black}{44.7\%}. Notably, through this comprehensive evaluation, we newly identify \textcolor{black}{32,995} crystal structures that are fully phonon-stable across the entire Brillouin zone \textcolor{black}{under the strict threshold of -0.001 THz}. These structures constitute a substantial and reliable pool of candidate materials for subsequent materials design and discovery, highlighting both the promise of generative models in expanding the known materials space and their current limitations in guaranteeing dynamical stability. \rev{The PhononBench web service, powered by the high-accuracy MatterSim potential, enables ultra-fast phonon predictions on a minute-level timescale and supports high-throughput evaluation of more than 10,000 materials per day, providing the technical foundation for the large-scale analysis performed in this work.}

\section*{Results}\label{sec-results}


\subsection*{Systematic Evaluation of Dynamical Stability in Crystal Generation Models}

\begin{figure*}[t!]
		\centering  
		\includegraphics[width=1.0\linewidth]{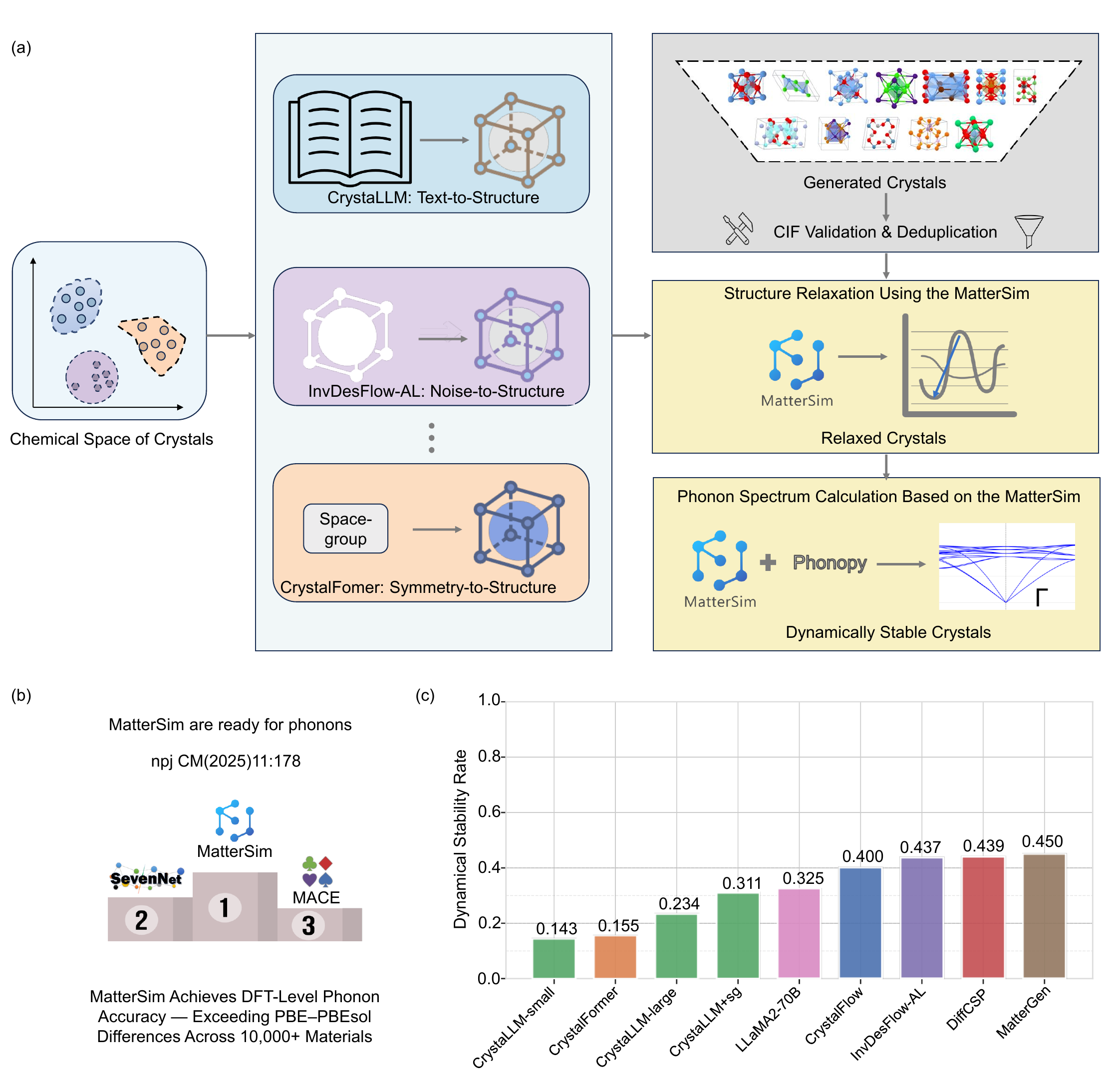}
		\caption{
\textbf{Systematic Evaluation of Dynamical Stability in Crystal Generation Models.}
(a) Workflow of this study. Eight generative models (CrystaLLM, MatterGen, DiffCSP, InvDesFlow-AL, CrystalFlow, CrystalFormer, etc.) were used to generate a total of 221,000 novel structures. After removing duplicates and post-processing for CIF validity, full phonon-spectrum calculations were performed using MatterSim combined with Phonopy on 108,843 successfully relaxed crystals. In total, \textcolor{black}{32,995} structures were found to be dynamically stable, corresponding to an overall stability ratio of 25.83\%.
(b) Based on the systematic phonon-spectrum evaluation of over 10,000 materials by Miguel A. L. Marques et al., we employed MatterSim-v1—which attains DFT-level accuracy with an average error smaller than the difference between PBE and PBEsol functionals—as the unified potential for all subsequent phonon calculations, ensuring consistency in evaluation standards.
(c) \textcolor{black}{Benchmark comparison of dynamical stability across crystal generation models on the MP20 dataset. The bar chart reports the fraction of dynamically stable structures under the phonon frequency threshold of $-0.1$~THz. GNN-based and diffusion-based models achieve consistently higher stability, with MatterGen (45.05\%), DiffCSP (43.94\%), and InvDesFlow-AL (43.78\%) ranking as the top-performing methods. In contrast, LLM/Transformer-based models show lower stability overall, with the best configuration (LLaMA2-70B) reaching 32.50\%.}
}
		\label{Benchmark-main} 
\end{figure*}

In this work, we employed \textcolor{black}{7} crystal generative models, including the LLM-based CrystaLLM~\cite{crystallm} and \textcolor{black}{LLaMA2-70B}~\cite{iclr2024LLM}; the graph-neural-network–enhanced diffusion models MatterGen~\cite{mattergen}, DiffCSP~\cite{diffcsp}, and InvDesFlow-AL~\cite{InvDesFlow-AL}; the flow-matching model CrystalFlow~\cite{CrystalFlow}; and the space-group-controlled CrystalFormer~\cite{crystalformer} (covering variants trained on \textcolor{black}{same} datasets)—to generate a total of \textcolor{black}{247,437} novel crystalline materials (Fig.~\ref{Benchmark-main}). After generation, we removed duplicates with respect to each model’s training set and performed post-processing to correct malformed CIF files. Structural relaxations were then carried out for the remaining crystals, of which \textcolor{black}{135,280} converged successfully. Subsequently, we conducted full phonon-spectrum calculations on the relaxed structures with MatterSim~\cite{mattersim} coupled to Phonopy~\cite{phonopy-phono3py-JPCM,phonopy-phono3py-JPSJ}, \textcolor{black}{identifying 32,995 crystals as dynamically stable under the strict criterion of $-0.001$~THz. Notably, the average dynamical stability across all models exhibits a strong dependence on the chosen phonon-frequency threshold, increasing substantially from 18.86\% at $-0.001$~THz to 19.57\% at $-0.01$~THz, 32.19\% at $-0.1$~THz, and 64.03\% at $-1.0$~THz. However, even under more relaxed thresholds, a large fraction of generated structures remain dynamically unstable.}Our systematic evaluation demonstrates that achieving dynamical stability remains a significant challenge for all current crystal generative models.

To ensure fair comparability across different generative models, we adopt the ratio of dynamically stable structures as a unified evaluation metric. Specifically, this ratio is defined as the number of phonon-stable crystals (i.e., those without imaginary phonon modes) divided by the number of successfully relaxed structures. This metric effectively eliminates biases arising from differences in novelty, CIF compliance rates, and relaxation success rates among models. Furthermore, our experiments show that the stability ratio produced by a generative model converges once the sample size exceeds approximately 4,000, with the remaining uncertainty being sufficiently small to avoid affecting the performance ranking reported in this study (Detailed convergence analysis of the dynamical stability rate is provided in Fig.~S.2 of the Supplementary Materials.). Except for CrystaLLM, all models in the figure exceed this convergence threshold in terms of phonon-calculated samples.

\textcolor{black}{As shown in Table~\ref{tab:stability_main}, we categorize the models into GNN-based and LLM/Transformer-based approaches, and further group them according to the training datasets (MP20 and Alex20). All generated structures are deduplicated against their corresponding training sets. The table reports dynamical stability under multiple phonon-frequency thresholds, providing a comprehensive evaluation across different criteria. For a more physically meaningful comparison, we focus on the threshold of $-0.1$~THz as a unified standard. As illustrated in Fig.~\ref{Benchmark-main}(c), which specifically presents the results at this threshold, the top three models on the MP20 dataset are MatterGen (\textcolor{black}{45.05\%}), DiffCSP (\textcolor{black}{43.94\%}), and InvDesFlow-AL (\textcolor{black}{43.78\%}), all of which belong to the GNN-based category and significantly outperform LLM-based models. This suggests that GNN-based approaches are more effective at capturing local geometric constraints and physical interactions, thereby facilitating the generation of dynamically stable structures. Notably, CrystalFlow demonstrates the highest generation efficiency among all models.}
In terms of model architecture, MatterGen employs a GemNet-based~\cite{GemNet} diffusion generative framework, whereas DiffCSP uses EGNN~\cite{egnn}. Both are diffusion-based models, with their performance ranking first and second, further highlighting the notable advantage of diffusion frameworks for crystal generation tasks. At the data representation level, MatterGen uses Cartesian coordinates (atomic positions), while the DiffCSP series employs fractional coordinates. This difference may affect model performance when scaling to supercells: Cartesian coordinates preserve the actual interatomic distances, whereas fractional coordinates change relative to the cell size, introducing a physical inconsistency that could contribute to performance differences. In addition, MatterGen utilizes D3PM~\cite{d3mp}, specifically designed for discrete data, which may also be an important factor in its superior performance. Detailed statistics on crystal generation, structural relaxation, and dynamical stability for all evaluated generative models are summarized in Fig.~S.1 and Table~S.1 in the Supplementary Materials.

\textcolor{black}{Furthermore, on the Alex20 dataset (Table~\ref{tab:stability_main}), all models exhibit improved stability, with MatterGen (57.89\%) and InvDesFlow-AL (52.65\%) showing particularly strong performance. This indicates that larger-scale or higher-quality training data can enhance the physical plausibility of generated structures (see Appendix for further analysis on data scaling). Overall, although both model type and dataset influence stability performance, GNN-based methods remain dominant in achieving dynamical stability under a unified threshold.} Moreover, CrystalFormer ranks third and achieves relatively good performance by constraining crystal generation in a manner that aligns better with physical intuition. However, this approach reduces novelty, as the strong constraints make it difficult to generate low-symmetry yet potentially stable materials. A detailed discussion is provided in Fig.~\ref{crystalfomer-gen}.

In this test, \textcolor{black}{CrystaLLM-small} achieved a dynamical stability ratio of only \textcolor{black}{14.3\%}, ranking last. It should be noted that although the model supports unconditional generation, the generative quality can degrade significantly in the absence of effective prompts due to the inherent next-token prediction mechanism of large language models. To ensure fair comparison, we retained its unconditional generation mode, producing a total of 16,000 crystals. After post-processing, only 2,074 structures were valid, of which 1,951 converged successfully, and ultimately only 58 were dynamically stable. \textcolor{black}{This observation motivates a further investigation into whether the performance of LLM-based generators can be improved with more advanced configurations. To further investigate the impact of model scale and prompting strategies, we additionally evaluated recent LLM-based crystal generation methods under more optimized settings. As shown in Table~\ref{tab:stability_main}, CrystaLLM-large models with prompt guidance exhibit substantially improved stability compared to the small, unprompted version. For instance, under the strictest phonon threshold of $-0.001$~THz, the stability increases from 3.0\% to 18.4\% and 19.6\% for the large models with formula-based prompts, surpassing CrystalFlow (16.8\%) and CrystalFormer (11.6\%) trained on MP20. Furthermore, the LLaMA2-70B~\cite{iclr2024LLM} model achieves an even higher stability of 21.7\%, and up to 23.9\% under certain sampling configurations, demonstrating that both prompt design and sampling strategies play a critical role in improving the quality of generated structures.}

\textcolor{black}{These results lead to a more nuanced understanding of LLM-based crystal generation. While naive, unprompted LLM models perform poorly in this task, their performance can be significantly improved through scaling, prompt engineering, and sampling optimization. In particular, large LLMs with carefully designed prompts can achieve stability levels comparable to, and in some cases exceeding, traditional GNN-based generators. However, it is important to interpret these improvements in the context of fundamental differences between the two paradigms. LLM-based approaches typically involve substantially larger models and rely heavily on prompt design and sampling strategies, introducing additional complexity and computational cost. In contrast, GNN-based generators are inherently tailored to atomistic systems and do not require external prompting, making them more straightforward and efficient for practical applications.}


\subsection*{Dynamical Stability Analysis of Space-Group-Constrained Crystal Generation}

\begin{figure*}[t!]
		\centering  
		\includegraphics[width=1.0\linewidth]{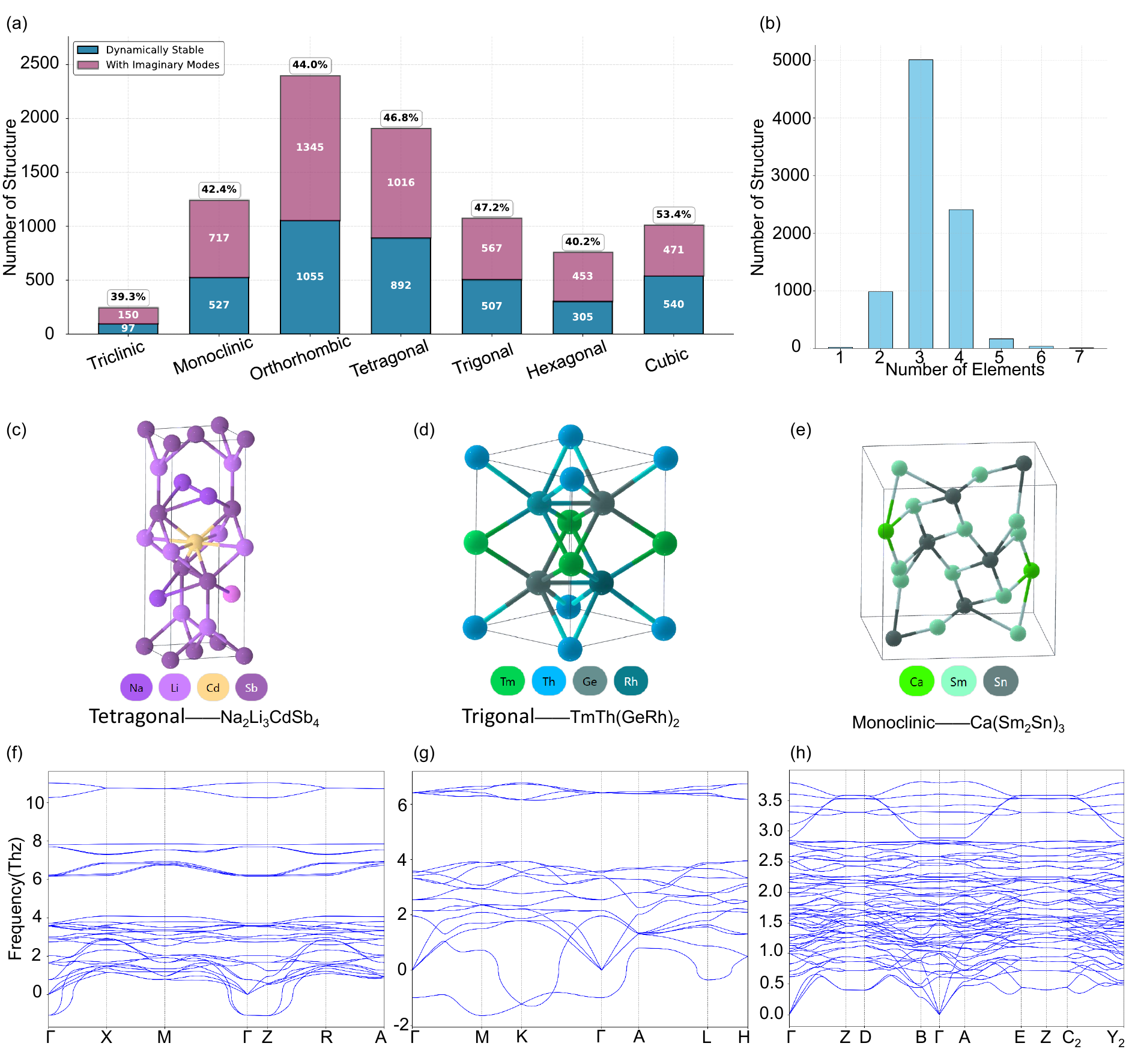}
		\caption{
\textbf{Dynamical Stability Analysis of Space-Group-Constrained Crystal Generation.}
(a) Distribution of the generated crystal structures across the seven crystal systems and their dynamical stability ratios (dark blue: dynamically stable; purple: containing imaginary modes). The cubic system exhibits the highest stability \textcolor{black}{(53.4\%)}, whereas the triclinic system shows the lowest \textcolor{black}{(39.3\%)}.
(b) Distribution of the number of elemental components in the generated materials, where ternary and quaternary compounds account for more than half of the dataset, consistent with the trend observed in Materials Project.
(c) Tetragonal structure: $\text{Na}_2\text{Li}_3\text{CdSb}$
(d) Trigonal structure: $\text{TmTh}(\text{GeRh})_2$
(e) Monoclinic structure: $\text{Ca}(\text{Sm}_2\text{Sn})_3$
(f)–(h) Phonon spectra corresponding to (c)–(e). $\text{Na}_2\text{Li}_3\text{CdSb}$ and $\text{TmTh}(\text{GeRh})_2$ exhibit pronounced imaginary phonon modes, indicating dynamical instability, whereas $\text{Ca}(\text{Sm}_2\text{Sn})_3$ shows no imaginary modes and is thus dynamically stable. These examples illustrate the ability of CrystalFormer to generate materials with complex chemistries, while also highlighting the remaining challenges in ensuring dynamical stability. They underscore the importance of incorporating explicit stability constraints or performing posterior stability screening within the generation pipeline.}
		\label{crystalfomer-gen} 
\end{figure*}


This section analyzes the dynamical stability of crystal structures generated by a space group constrained generative model. We employed CrystalFormer~\cite{crystalformer} (Alex20) to generate a total of 40,000 crystal structures, with their space-group distribution matched to that of the Alex20 training set. \textcolor{black}{After removing structures overlapping with Alex20, 8,986 unique crystals were retained for further relaxation and phonon calculations, among which 8,642 successfully completed the phonon calculations. As shown in Fig.~\ref{crystalfomer-gen}(a), the generated crystals exhibit a distribution across the seven crystal systems that is highly consistent with the Materials Project dataset. Among the 8,642 structures with successful phonon calculations, 3,923 structures (45.39\%) are dynamically stable under a phonon frequency threshold of $-0.1$~THz.} The elemental composition distribution (Fig.~\ref{crystalfomer-gen}(b)) shows that ternary and quaternary compounds account for more than half of the generated materials, in agreement with the statistical trends observed in the Materials Project. \textcolor{black}{The dynamical stability ratio varies considerably among different crystal systems: cubic structures show the highest stability ratio (53.4\%), followed by trigonal (47.2\%), tetragonal (46.8\%), orthorhombic (44.0\%), monoclinic (42.4\%), and hexagonal (40.2\%), while triclinic structures exhibit the lowest stability ratio at only 39.3\%.} These results suggest a possible correlation between structural symmetry and dynamical stability: high-symmetry crystal systems such as cubic tend to exhibit higher stability ratios, whereas low-symmetry systems such as triclinic show markedly reduced stability. \textcolor{black}{This trend can be understood from symmetry constraints imposed by Neumann’s principle~\cite{nye1985physical, tinkham2003group,born1996dynamical}, which require the force constant (Hessian) matrix to remain invariant under all symmetry operations of the crystal, thereby reducing the number of independent degrees of freedom and restricting the admissible eigenvalue spectrum. As a result, higher symmetry suppresses the emergence of soft modes, particularly negative eigenvalues associated with imaginary phonon frequencies. The consistency of this symmetry–stability relationship in AI-generated structures further indicates that the generative model captures fundamental physical constraints governing lattice dynamics.}  Moreover, the elemental distribution of the generated materials closely matches that of the real dataset, indicating that the model achieves good coverage of chemical compositional diversity. Nevertheless, there remains room for improvement in the dynamical stability of the generated structures. It is also noteworthy that, relative to generative models without space-group constraints—such as InvDesFlow-AL~\cite{InvDesFlow-AL} and DiffCSP~\cite{diffcsp}—the space-group–restricted model exhibits significantly reduced novelty, reflecting the inherent limitations imposed by symmetry constraints on the generative space.

Figures~\ref{crystalfomer-gen}(c)–(e) present three representative crystal structures generated by CrystalFormer: (c) a tetragonal $\text{Na}_2\text{Li}_3\text{CdSb}$ compound, (d) a trigonal $\text{TmTh}(\text{GeRh})_2$ compound, and (e) a monoclinic $\text{Ca}(\text{Sm}_2\text{Sn})_3$ compound. Their corresponding phonon spectra are shown in Figures~\ref{crystalfomer-gen}(f)–(h). The phonon results indicate that both $\text{Na}_2\text{Li}_3\text{CdSb}$ and $\text{TmTh}(\text{GeRh})_2$ exhibit pronounced imaginary frequencies across multiple phonon branches, signaling strong dynamical instabilities and suggesting that these structures may reside near saddle points or shallow extrema on the potential energy landscape. In contrast, the phonon spectrum of $\text{Ca}(\text{Sm}_2\text{Sn})_3$ shows no imaginary modes throughout the Brillouin zone, demonstrating robust dynamical stability. The extensive imaginary modes observed in the first two structures may originate from unfavorable bonding configurations, such as excessively short interatomic distances leading to strong repulsive interactions, or local coordination environments that cannot be dynamically sustained. These case studies clearly illustrate the dual nature of the current generative model: while it can construct chemically complex structures that relax to energetically reasonable configurations, many of these structures still face significant dynamical challenges. This highlights that energy-based criteria alone are insufficient for crystal generation tasks and underscores the necessity of incorporating explicit dynamical stability constraints or performing systematic post-generation screening.

\subsection*{Dynamical Stability Analysis of Property-Constrained Crystal Generation}

\begin{figure*}[t!]
		\centering  
		\includegraphics[width=1.0\linewidth]{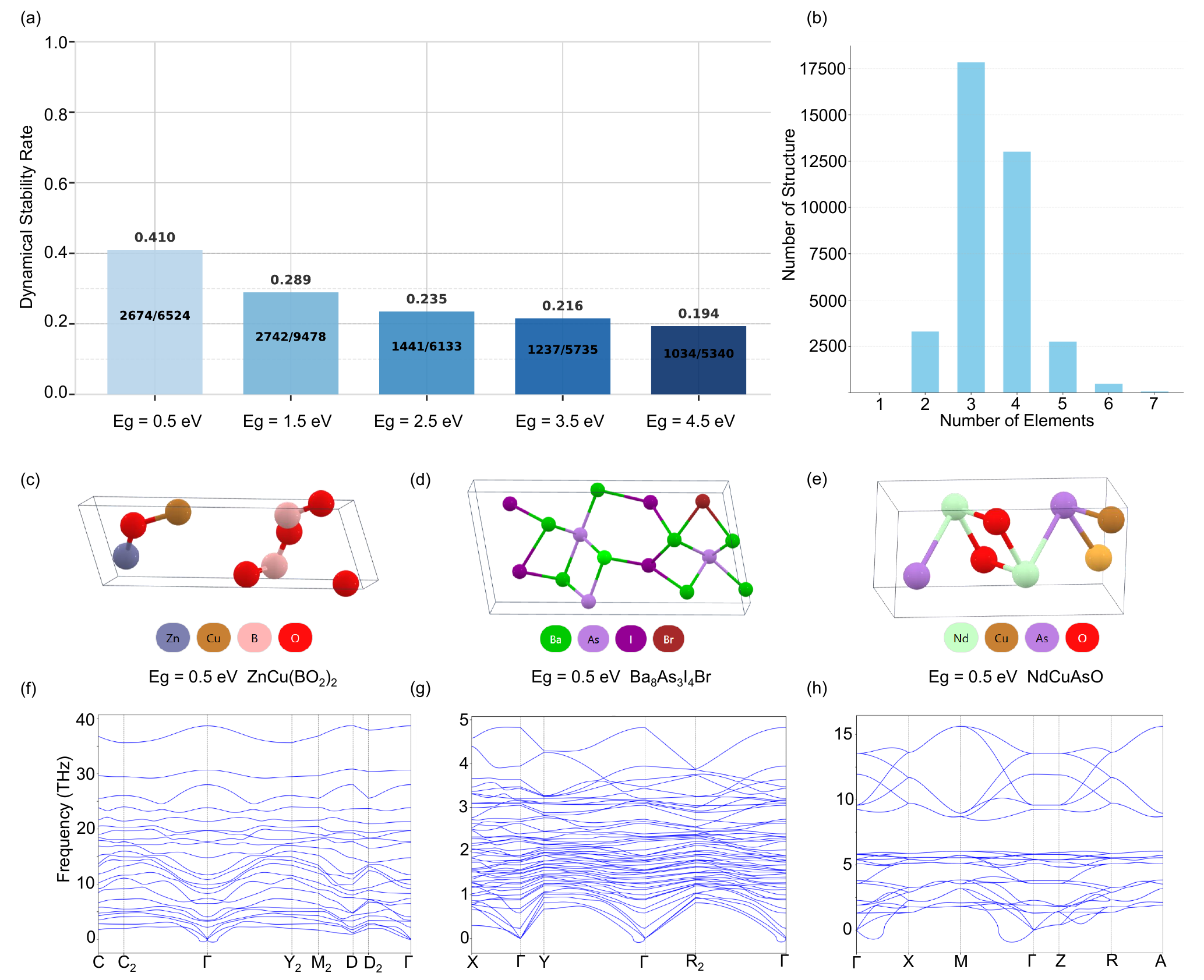}
		\caption{
\textbf{Dynamical Stability Analysis of Property-Constrained Crystal Generation.}
(a) Dynamical stability ratios of generated materials under different band-gap constraints. Among the 33,210 structures subjected to phonon analysis, the $E_g = 4.5$ eV condition yields the lowest stability, with \textcolor{black}{1,037} dynamically stable structures out of 5,340 samples \textcolor{black}{(19.4\%)}, whereas $E_g = 0.5$ eV yields the highest stability, with \textcolor{black}{2,674} stable structures among 6,524 samples \textcolor{black}{(41.0\%)}. The stability ratios for the other settings are \textcolor{black}{28.9\%} for $E_g = 1.5$ eV (\textcolor{black}{2,742}/9,478), and \textcolor{black}{23.5\%} for both $E_g = 2.5$ eV (\textcolor{black}{1,441}/6,133) and $E_g = 3.5$ eV (\textcolor{black}{1,237}/5,735). Overall, the stability rate under band-gap conditioning remains relatively low, reaching only \textcolor{black}{26.8\%}, implying that subsequent phonon validation with QE or VASP would incur considerable computational cost and pose challenges for large-scale applications.
(b) Distribution of the number of elemental components in the generated materials.
(c)–(e) Three representative crystal structures generated by MatterGen under the $E_g = 0.5$ eV constraint: ZnCu(BO$_2$)$_2$, Ba$_8$As$_3$I$_4$Br, and NdCuAsO.
(f)–(h) Phonon spectra corresponding to panels (c)–(e). All three structures exhibit pronounced imaginary (negative) frequencies across multiple phonon branches, indicating strong dynamical instabilities in their current configurations.
}
		\label{Figure-3-bandgap} 
\end{figure*}

This section presents functional materials generation using the band-gap--conditioned MatterGen model. To ensure statistical significance, we predetermined that each band-gap condition should yield at least $\sim$4,000 unique crystals for stability evaluation. In total, 56,000 crystals were generated: 16,000 samples for $E_g = 1.5\ \text{eV}$, and 10,000 samples each for $E_g = 0.5\ \text{eV}$, $2.5\ \text{eV}$, $3.5\ \text{eV}$, and $4.5\ \text{eV}$. After removing structures overlapping with the MP20 training set, 33,210 crystals were subjected to phonon calculations to assess their dynamical stability. As shown in Fig.~\ref{Figure-3-bandgap}(a), the dynamical stability exhibits a clear dependence on the target band-gap conditions. The $E_g = 4.5\ \text{eV}$ condition shows the lowest stability: only \textcolor{black}{1,037} out of 5,340 evaluated samples are dynamically stable \textcolor{black}{(19.4\%)}. In contrast, the $E_g = 0.5\ \text{eV}$ condition yields the highest stability, with \textcolor{black}{2,674} stable structures among 6,524 samples \textcolor{black}{(41.0\%)}. The stability ratios for the remaining conditions are as follows: \textcolor{black}{28.9\%} for $E_g = 1.5\ \text{eV}$ (\textcolor{black}{2,742} stable out of 9,478), and \textcolor{black}{23.5\%} for both $E_g = 2.5\ \text{eV}$ and $3.5\ \text{eV}$ (\textcolor{black}{1,441} stable out of 6,133 and \textcolor{black}{1,237} stable out of 5,735, respectively). Overall, the dynamical stability in band-gap--conditioned functional materials generation remains relatively low. Even with the advanced fine-tuned MatterGen framework, the overall stability ratio reaches only \textcolor{black}{26.8\%}. This indicates that relying on conventional first-principles packages such as Quantum~ESPRESSO (QE) or the Vienna \textit{Ab initio} Simulation Package (VASP) for subsequent phonon calculations would incur substantial computational cost, posing challenges for practical large-scale deployment.

From the elemental composition distribution (Fig.~\ref{Figure-3-bandgap}(b)), we observe that, under the band-gap constraint, ternary and quaternary compounds account for more than 50\% of the generated materials. This trend closely aligns with the statistical patterns found in the Materials Project database, indicating that the model maintains a reasonable exploration of chemical space in controlled-generation tasks. In addition, Fig.~\ref{Figure-3-bandgap}(c)–(e) illustrates three representative crystal structures generated by MatterGen---ZnCu(BO$_2$)$_2$, Ba$_8$As$_3$I$_4$Br, and NdCuAsO---with a target band gap of $E_g = 0.5~\text{eV}$. Their corresponding phonon spectra are shown in Fig.~\ref{Figure-3-bandgap}(f)–(h). The phonon calculations reveal pronounced imaginary (negative) frequencies across multiple phonon branches for all three materials, indicating strong dynamical instabilities in their current structural configurations. To facilitate further analyses of structural reliability and potential properties, we will release all phonon calculations and optimized structures for the full set of 33,210 generated materials, providing a comprehensive benchmark for subsequent validation and methodological developments. The statistics of structural relaxation and dynamical stability, stratified by different band-gap ranges, are summarized in Table S.2 of the Supplementary Materials.

\subsection*{\textcolor{black}{Physical Origins of Dynamical Instability}}
\begin{figure*}[t!]
		\centering  
		\includegraphics[width=1.0\linewidth]{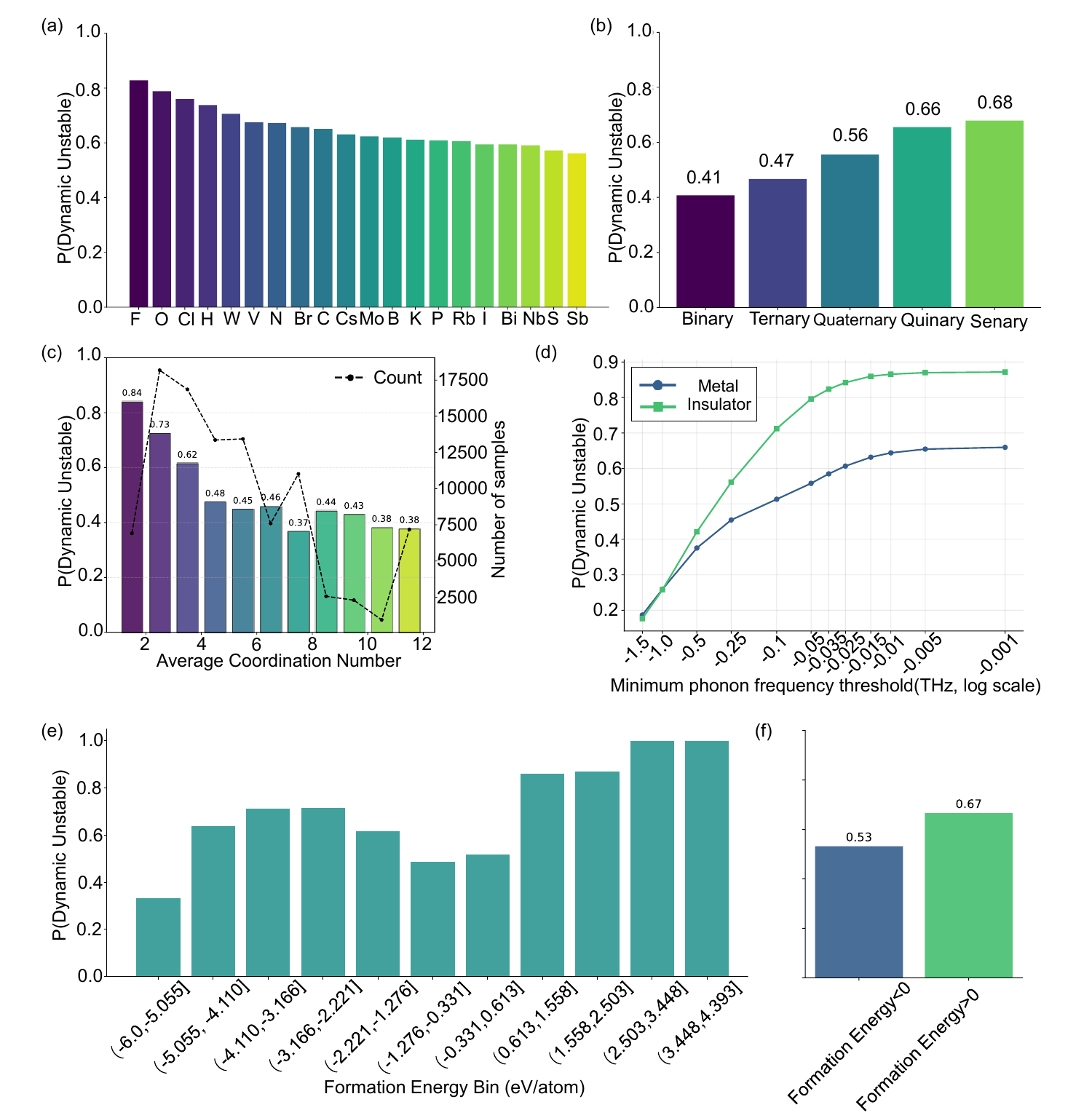}
		\caption{
\textcolor{black}{\textbf{Physical Origins of Dynamical Instability.}
(a) Element-resolved instability probability under an imaginary-frequency threshold of $-0.1$~THz. (b) Instability probability as a function of chemical complexity, measured by the number of constituent elements. (c) Instability probability versus average coordination number (CN).(d) Comparison of instability behavior between metallic and insulating systems. (e) Relationship between thermodynamic formation energy.}}
		\label{Figure-phy-ins} 
\end{figure*}

\textcolor{black}{To uncover the physical origins of dynamical instability in AI-generated structures, we perform a systematic analysis from multiple complementary perspectives, as summarized in Fig.~\ref{Figure-phy-ins}. Unless otherwise specified, all results are reported using a representative imaginary-frequency threshold of $-0.1$~THz. Fig.~\ref{Figure-phy-ins}(a) presents an element-resolved analysis of instability probability. A clear chemical dependence is observed: structures containing highly electronegative nonmetals such as F, O, Cl, H, Br, and I exhibit the highest instability probabilities (exceeding $\sim 65\%$), followed by alkali metals (Cs, K, Rb) and selected transition metals (V, Mo, W). This trend indicates that dynamical instability is not randomly distributed but strongly governed by elemental chemistry. Physically, directional and rigid bonding associated with electronegative elements makes structures highly sensitive to small geometric distortions, while large ionic radii and weak bonding in alkali metals lead to lattice softening. Transition metals further introduce complexity through variable coordination environments and $d$-electron bonding, increasing susceptibility to unstable modes.}

\textcolor{black}{Fig.~\ref{Figure-phy-ins}(b) examines the role of chemical complexity by grouping structures according to the number of constituent elements. A monotonic increase in instability probability is observed with increasing compositional complexity. This behavior reflects the rapid expansion of configurational space in multicomponent systems, where satisfying local bonding preferences, charge balance, and geometric constraints simultaneously becomes increasingly difficult. As a result, local distortions and lattice frustration emerge more readily, promoting dynamical instability.
Fig.~\ref{Figure-phy-ins}(c) shows the dependence of dynamical instability on the average coordination number (Coordination Number, CN), which characterizes local atomic connectivity. A pronounced inverse correlation is observed: the instability probability decreases from approximately $84\%$ in the low-coordination regime (average CN $\approx 1.5$) to about $45\%$ at intermediate coordination (average CN $\approx 5.5$), and further to $\sim 38\%$ at high coordination (average CN $\gtrsim 9$). This trend indicates that under-coordinated structures, lacking sufficient geometric constraints, are more prone to soft phonon modes. In contrast, as the coordination number increases, atoms become more tightly packed and effectively confined within local potential wells, leading to enhanced lattice connectivity and rigidity, which stabilizes the structure. High-coordination environments are commonly associated with densely packed structures, which tend to exhibit enhanced mechanical rigidity.}

\textcolor{black}{Fig.~\ref{Figure-phy-ins}(d) compares instability behavior between metallic and insulating systems. At moderate instability levels, insulating structures exhibit a higher fraction of unstable cases than metallic ones, whereas the difference diminishes for stronger instabilities. This suggests that weak instabilities are often associated with subtle structural distortions—such as symmetry breaking or octahedral tilting—which are more prevalent in insulating systems, while strong instabilities are primarily driven by geometrically unrealistic configurations and affect both classes similarly.
Figures.~\ref{Figure-phy-ins}(e)-(f) examine the relationship between thermodynamic and dynamical stability. From a physical perspective, these two notions describe complementary aspects of stability: thermodynamic stability reflects the global energetic favorability against decomposition, whereas dynamical stability is governed by the local curvature of the potential energy surface. Consistent with this picture, structures with positive formation energy exhibit a higher probability of dynamical instability (66.62\%) compared to those with negative formation energy (53.21\%). However, the overall correlation remains weak, indicating that thermodynamic instability contributes to, but does not dominate, dynamical instability. This observation suggests an intrinsic decoupling between global energetic ordering and local structural stability, implying that phonon stability is primarily governed by local structural factors rather than global thermodynamic considerations. In this context, it is worth noting that benchmarks such as MatBench-Discovery~\cite{matbench} focus primarily on thermodynamic stability, whereas our benchmark explicitly targets dynamical (phonon) stability, highlighting a fundamentally different and complementary aspect of materials assessment.}

\textcolor{black}{Taken together, these results reveal that dynamical instability in AI-generated structures has clear physical origins, arising from the interplay of chemical composition, structural complexity, local coordination, and electronic character. Rather than being random prediction errors, the observed instabilities reflect systematic challenges of current generative models in capturing strong bonding constraints, lattice flexibility, and complex coordination environments. More detailed analysis is provided in the Supplementary Materials.}

\subsection*{Dynamically Stable Crystal Structures}

\begin{figure*}[t!]
		\centering  
		\includegraphics[width=1.0\linewidth]{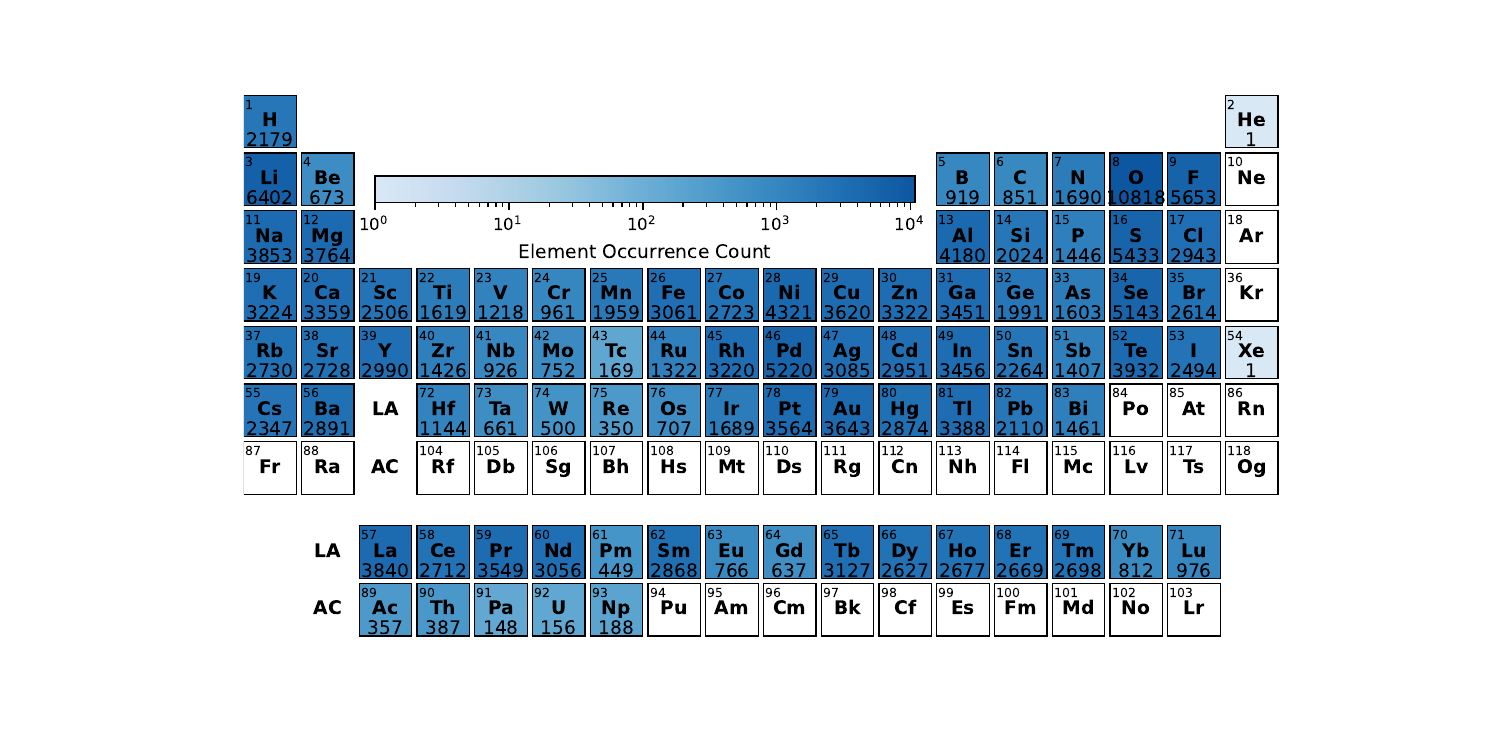}
		\caption{
\textbf{Elemental distribution heatmap of dynamically stable crystals.}
The heatmap illustrates the distribution of chemical elements in \textcolor{black}{32,995} newly discovered crystal structures with phonon (dynamical) stability, generated by the crystal generation model. Each element is colored according to its occurrence frequency in the stable crystal set, with darker colors indicating higher frequencies. The analysis shows that oxygen (O) is the most prevalent element (\textcolor{black}{10,818} occurrences), followed by lithium (Li, \textcolor{black}{6,402}) and fluorine (F, \textcolor{black}{5653}), whereas noble gas elements appear only rarely. This distribution clearly indicates that current generative models tend to predict stable compounds containing chemically active elements, in good agreement with established chemical intuition.
}
		\label{Figure-4} 
\end{figure*}

\begin{figure*}[t!]
		\centering  
		\includegraphics[width=1.0\linewidth]{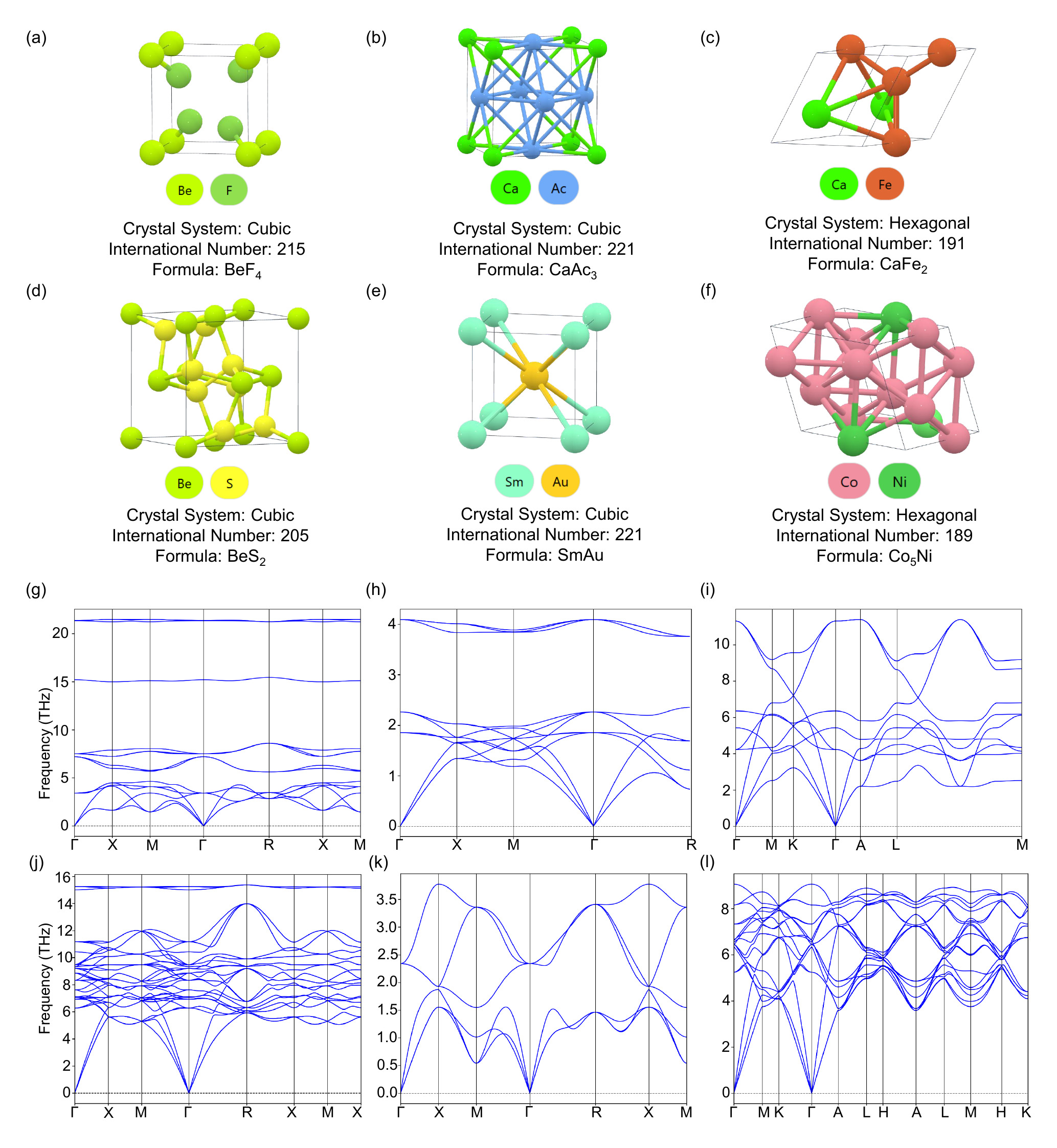}
		\caption{
\textcolor{black}{\textbf{Representative examples of AI-generated dynamically stable crystal structures.}
Panels (a)-(f) display the crystal structures of selected materials (BeF$_4$, CaAc$_3$, CaFe$_2$, Co$_5$Ni, SmAu, etc.), covering cubic and hexagonal crystal systems (with space group numbers 215, 221, 191, 189, etc.). Panels (g)-(l) show the corresponding phonon spectra obtained from DFT calculations. All structures exhibit no imaginary phonon frequencies throughout the entire Brillouin zone, indicating good dynamical stability.}}
		\label{Figure-dft} 
\end{figure*}

In this study, we conducted a systematic evaluation of the dynamical stability of generated crystal materials and identified \textcolor{black}{32,995} dynamically stable crystal structures. This discovery provides a rich pool of candidate systems for materials science research and substantially expands the database of known stable materials. Although current crystal generation models are not explicitly optimized for phonon stability during training, they demonstrate a capability for novel materials design that significantly surpasses traditional manual design approaches. Through elemental statistical analysis (Fig.~\ref{Figure-4}), we observe clear regularities in the elemental distribution of these stable crystals. Oxygen (O) appears most frequently, with \textcolor{black}{10,818} occurrences, followed by lithium (Li) with \textcolor{black}{6,402} occurrences, and fluorine (F) with \textcolor{black}{5653} occurrences. In contrast, noble gas elements are rarely present in dynamically stable crystals, consistent with their chemical inertness. These trends indicate that the generative model effectively captures fundamental chemical principles governing real materials. All evaluated crystal structures and their corresponding phonon calculation details will be fully open-sourced, providing a transparent and verifiable data foundation for the community. The validated dynamically stable crystals constitute a reliable candidate set for exploring novel functional materials, enabling density functional theory practitioners to focus on functional property investigations without concerns about basic dynamical stability. The open availability of this dataset is expected to accelerate the discovery of functional materials and advance computational materials science toward higher accuracy and efficiency. The elemental distribution of the \textcolor{black}{32,995} dynamically stable crystals is shown in Fig.~S.3 of the Supplementary Materials.

\textcolor{black}{In addition, to validate the reliability of phonon predictions based on MatterSim, we constructed an independent validation set consisting of 120 structures and performed density functional theory (DFT) phonon calculations. Based on this dataset, a systematic comparison among three representative phonon prediction models was conducted, with results summarized in Table~\ref{dft-120-mul-model}. Under the strictest criterion of $-0.001$~THz, MatterSim achieves the best performance in both true stable rate (TS = 97.62\%) and overall accuracy (Acc = 88.07\%), while its false unstable rate (FU = 2.38\%) is significantly lower than those of other models, indicating high reliability in identifying stable structures. In contrast, PET-OAM-XL exhibits extremely high true unstable (TU) rates but markedly low TS, revealing a strong systematic bias toward overpredicting instability. The eSEN-30M-OAM model shows a more balanced performance between stable and unstable classifications, yet still demonstrates a moderate conservative bias, tending to misclassify stable structures as unstable. This trend remains consistent across different frequency thresholds, indicating that MatterSim provides a more balanced and reliable classification of dynamical stability without exhibiting significant systematic bias.}
\textcolor{black}{To validate the reliability of phonon predictions based on MatterSim, we constructed an independent validation set consisting of 120 structures by randomly sampling across different stability categories and diverse chemical spaces, and performed density functional theory (DFT) phonon calculations. From this set, representative dynamically stable structures are presented in Figures.~\ref{Figure-dft}(a)–(f) show the corresponding crystal structures, covering cubic and hexagonal systems with diverse elemental compositions, while (g)–(l) display the associated DFT phonon spectra. All structures exhibit no imaginary frequencies throughout the Brillouin zone, confirming their dynamical stability. The complete validation results are provided in the Supplementary Materials.}
\subsection*{Summary of Model Architectures and Inference Performance}

\begin{table}[htbp]
\small
\centering
\color{black}
\caption{\textcolor{black}{Normalized confusion matrix (\%) for predicted dynamical stability.
TS: true stable, FU: false unstable, TU: true unstable, FS: false stable.
The Benchmark of Ref. ~\cite{PhonnBench} contains 8189 stable and 1769 unstable structures.
The subset contains 50 stable and 50 unstable structures.$\clubsuit$ denotes additional tests introduced in this work.}}
\begin{tabular}{llcccc}
\toprule
Dataset & Model & TS$\uparrow$ (\%) & FU$\downarrow$ (\%) & TU$\uparrow$ (\%) & FS$\downarrow$ (\%) \\
\midrule
\multirow{9}{*}{Benchmark of Ref.~\cite{PhonnBench}}
 & M3GNet & 87 & 13 & 73 & 27 \\
 & CHGNet & 77 & 23 & 73 & 27 \\
 & MACE-MP-0 & \textbf{95} & 5 & 73 & 27 \\
 & SevenNet-0 & 81 & 19 & 80 & 20 \\
 & ORB & 15 & 85 & 92 & 8 \\
 & eqV2-M & 7 & 93 & \textbf{94} & \textbf{6} \\
  & PBESol (DFT reference) & 97 & 3 &  76 & 24 \\
 & MatterSim-v1 & \textbf{95} & \textbf{5} & 75 & 25 \\
 & PET-OAM-XL$\clubsuit$ & 63 & 37 & 91 & 9 \\
\midrule
\multirow{4}{*}{Subset (100)}
 & eSEN-30M-OAM$\clubsuit$ & 76 & 24 & 84 & 16 \\
  & PET-OAM-XL$\clubsuit$ & 68 & 32 & \textbf{98} & \textbf{2} \\
   & PBESol (DFT reference)$\clubsuit$ & 100 & 0 &  80 & 20 \\
 & MatterSim-v1$\clubsuit$ & \textbf{96} & \textbf{4} & 80 & 20 \\
\bottomrule
\end{tabular}
\label{tab:phonon_confusion}
\end{table}

\begin{table}[htbp]
\centering
\color{black}
\small
\caption{\textcolor{black}{Comparison of dynamical stability prediction under different frequency thresholds on a validation subset of 120 AI-generated structures, benchmarked against DFT phonon calculations. TS: true stable, FU: false unstable, TU: true unstable, FS: false stable. Acc: overall accuracy. Arrows indicate whether higher ($\uparrow$) or lower ($\downarrow$) values are better.}}
\begin{tabular}{c|c|ccccc}
\hline
Threshold (THz) & Method 
& TS (\%) $\uparrow$ 
& FU (\%) $\downarrow$ 
& TU (\%) $\uparrow$ 
& FS (\%) $\downarrow$ 
& Acc (\%) $\uparrow$ \\
\hline

\multirow{3}{*}{-0.001}
&PET-OAM-XL   
& 30.95 & 69.05 & \textbf{97.01} & \textbf{2.99}  & 71.56 \\
& eSEN-30M-OAM 
& 83.33 & 16.67 & 88.06 & 11.94  & 86.24 \\
& MatterSim-v1    
& \textbf{97.62} & \textbf{2.38} & 74.63 & 25.37  & \textbf{88.07} \\

\hline

\multirow{3}{*}{-0.01}
&PET-OAM-XL   
& 30.95 & 69.05 & \textbf{97.01} & \textbf{2.99}  & 71.56 \\
& eSEN-30M-OAM 
& 83.33 & 16.67 & 88.06 & 11.94  & 86.24 \\
& MatterSim-v1    
& \textbf{97.62} &\textbf{2.38} & 74.63 & 25.37  & \textbf{88.07} \\

\hline

\multirow{3}{*}{-0.1}
&PET-OAM-XL   
& 30.23 & 69.77 & \textbf{96.97} & \textbf{3.03}  & 70.64 \\
& eSEN-30M-OAM 
& 88.37 & 11.63 & 89.39 & 10.61  & 88.99 \\
& MatterSim-v1    
& \textbf{95.35} & \textbf{4.65} & 74.24 & 25.76  & \textbf{89.91} \\

\hline

\multirow{3}{*}{-1.0}
&PET-OAM-XL   
& 52.00 & 48.00 & \textbf{94.92} & \textbf{5.08}  & 75.23 \\
& eSEN-30M-OAM 
& 90.00 & 10.00 & 81.36 & 18.64  & 85.32 \\
& MatterSim-v1    
& \textbf{92.00} & \textbf{8.00} & 79.66 & 20.34  & \textbf{85.32} \\

\hline
\end{tabular}
\label{dft-120-mul-model}
\end{table}

\begin{table}[h]
\small
\centering
\color{black}
\caption{\textcolor{black}{Inference efficiency comparison for phonon calculations on 100 structures.}}
\begin{tabular}{lccc}
\hline
Model & Total Time (100 structures) & Avg Time / Structure  \\
\hline
PET-OAM-XL & 7 h 33 m 13 s & 271.9 s \\
eSEN-30M-OAM & 27 h 31 m 59 s & 990.0 s \\
MatterSim-v1 & 3 h 59 m 36 s & \textbf{143.8 s}  \\
\hline
\end{tabular}
\label{tab:inference_speed}
\end{table}

\begin{table}[htbp]
\centering
\color{black}
\caption{\textcolor{black}{Dynamical stability of different crystal generation models evaluated by phonon calculations. 
Stability is defined as the fraction of structures with minimum phonon frequency above given thresholds (THz). 
CrystaLLM (small) denotes a small model without prompting. CrystaLLM (large) uses chemical formula prompts, 
while CrystaLLM (+sg) additionally incorporates space group information. LLaMA2-70B corresponds to a large model with an optimized prompting strategy.}}
\begin{tabular}{lcccc}
\hline
\multirow{2}{*}{Model} & \multicolumn{4}{c}{Frequency Threshold (THz)} \\
\cline{2-5}
 & $>-0.001$ & $>-0.01$ & $>-0.1$ & $>-1.0$ \\
\hline
\multicolumn{5}{c}{\textbf{GNN-based Generators}} \\
\hline
CrystalFormer (MP20)   & 0.1157 & 0.1198 & 0.1554 & 0.3058 \\
CrystalFlow (MP20)     & 0.1676 & 0.1763 & 0.4007 & 0.8260 \\
InvDesFlow-AL (MP20) & 0.2702 & 0.2778 & 0.4378  & 0.8037\\
DiffCSP (MP20)         & 0.2715 & 0.2791 & 0.4394 & 0.8030 \\
MatterGen (MP20)       & 0.2455 & 0.2557 & 0.4505 & 0.7915 \\
\hline
CrystalFormer (Alex20) & 0.3436 & 0.3555 & 0.4539 & 0.7234 \\
InvDesFlow-AL (Alex20)  & 0.3119 & 0.3403 & 0.5265 & 0.8775 \\
MatterGen (Alex20)     & 0.4099 & 0.4295 & 0.5789 & 0.8818 \\
\hline
\multicolumn{5}{c}{\textbf{LLM / Transformer-based Generators}} \\
\hline
CrystaLLM (small,MP20)     & 0.0297 & 0.0369 & 0.1435 & 0.5864 \\
CrystaLLM (large,MP20)     & 0.1840 & 0.1880 & 0.2340 & 0.4750 \\
CrystaLLM (+sg,MP20)       & 0.1960 & 0.2030 & 0.3110 & 0.5930 \\
LLaMA2-70B (opt,MP20)            & 0.2170 & 0.2250 & 0.3250 & 0.5780 \\
\hline
\end{tabular}
\label{tab:stability_main}
\end{table}

\begin{table}[h]
\centering
\small
\caption{Comparison of inference throughput for different generative models. To account for differences in maximum batch size due to model architecture and memory usage, speeds are normalized to an equivalent batch size of 200, enabling a fair and consistent evaluation of computational efficiency. The last column indicates the normalized generation speed in crystals per minute.}
\label{inferspeed}
\begin{tabular}{lcccc}
\toprule
\textbf{Model Name} & \textbf{Generated Items}  & \textbf{Total Time (h)} & \textbf{Speed (items/min)} & \textbf{Eq. Speed @200} $\mathbf{\uparrow}$
\\
\midrule
CrystalFlow~\cite{CrystalFlow}      & 16,000  & 0.64 & 416.6 & 333.3 \\
InvDesFlow-AL~\cite{InvDesFlow-AL}    & 25,000 & 1.7  & 240.0 & 48.0  \\
DiffCSP~\cite{diffcsp}          & 16,000   & 2.3  & 113.5 & 45.4  \\
CrystaLLM~\cite{crystallm}       & 30,000   & 9.4  & 52.9  & 42.3  \\
MatterGen~\cite{mattergen}        & 16,000   & 20.2 & 13.2  & 13.2  \\
CrystalFormer~\cite{crystalformer}    & 20,000  & 44.3 & 6.0   & 12.0  \\
\bottomrule
\end{tabular}
\end{table}

This section presents a systematic comparison of generation speed across different models. As shown in Table~\ref{paranum}, we report each model’s throughput for a single generation task (note: “Generated Items” refers to the number of structures used in this speed test, not the final number used for phonon calculations). To ensure fair comparison across architectures and memory usage, the generation speeds are normalized to an equivalent batch size of 200. All generation tasks, except for CrystalFormer, were performed on a single NVIDIA RTX 4090 GPU. The normalized inference throughput clearly reveals substantial differences in generation efficiency among the models. CrystalFlow demonstrates the highest performance, with an equivalent generation speed of 333.3 crystals per minute, significantly outperforming all other models. InvDesFlow-AL and DiffCSP form the second tier, with normalized speeds of 48.0 and 45.4 crystals per minute, respectively. CrystaLLM achieves 42.3 crystals per minute. MatterGen and CrystalFormer exhibit the lowest throughput, with equivalent speeds of only 13.2 and 12.0 crystals per minute; the slower speed of CrystalFormer is likely due to the use of its CPU version in our tests. In summary, while maintaining both novelty and stability of the generated materials, CrystalFlow demonstrates outstanding generation efficiency, exceeding the slowest model by more than an order of magnitude. This superior throughput provides a critical advantage for large-scale virtual screening and iterative optimization of crystal structures using generative models.


\begin{table}[h]
\centering
\small
\caption{Overview of the model architectures and their parameter counts. The compared models include both GNN-based crystal generative models and Transformer-based sequence models, covering a wide range of model capacities from lightweight (4.8M parameters) to large-scale (53.7M parameters).}
\label{paranum}
\begin{tabular}{lccc}
\hline
\textbf{Model Name} & \textbf{Training Set}& \textbf{Architecture Type}  & \textbf{Parameters} $\mathbf{\downarrow}$\\
\hline
CrystalFormer       & MP20      & Transformer & 4.8M \\
CrystalFormer       & Alex20      & Transformer & 4.8M \\
InvDesFlow-AL  & Alex20              & GNN        & 12.3M \\
DiffCSP      & MP20               & GNN        & 12.3M \\
CrystalFlow   & MP20             & GNN        & 20.9M \\
CrystaLLM  & MP20                 & Transformer & 26.1M \\
MatterGen   & MP20   & GNN        & 53.7M \\
MatterGen & Alex20  & GNN        & 53.7M \\
\hline
\end{tabular}
\end{table}

As shown in Table~\ref{paranum}, the compared models encompass two mainstream architectures: GNN-based crystal generative models and Transformer-based sequence models, with parameter counts ranging widely from lightweight 4.8M to large-scale 53.7M, reflecting diverse model capacities and design philosophies. In terms of architectural distribution, GNNs dominate, consistent with their natural suitability for modeling the periodic graph structures of crystals. Among them, MatterGen, with 53.7M parameters, is the largest model in this comparison, reflecting its design focus on high expressive capacity. CrystalFlow, another representative GNN model, has a moderate size of 20.9M parameters, while InvDesFlow-AL and DiffCSP share a compact GNN architecture with 12.3M parameters each. Regarding Transformer-based architectures, CrystaLLM (Small) is a 26.1M-parameter sequence model, exceeding the size of most GNN models. In contrast, CrystalFormer, also a Transformer, is the most lightweight model in this comparison, with only 4.8M parameters.


\label{csp-task}

\section*{Discussion}\label{sec12}
In practical applications of crystal generative models, different architectures exhibit notable differences in terms of novel material discovery and computational friendliness. For instance, the symmetry-constrained CrystalFormer shows a significant reduction in novelty: in one generation experiment, although 20,000 crystals were produced, only about 5,000 unique chemical formulas were obtained. This indicates that exploring new chemical spaces with such models requires generating a substantially larger number of samples to obtain sufficient novel materials. In contrast, the large language model-based CrystaLLM-small frequently encounters file parsing failures during generation, with failure rates reaching up to 90\% in severe cases, likely related to prompt design, highlighting practical limitations of this approach. \textcolor{black}{However, with improved prompting strategies and larger model scales, the performance of LLM-based generators can be significantly enhanced. In particular, CrystaLLM-large and LLaMA2-70B achieve substantially higher dynamical stability, reaching up to \textcolor{black}{32.5\%}, which even surpasses CrystalFormer (\textcolor{black}{15.5\%}) under the same setting.} Conventional equivariant graph neural network models such as MatterGen, DiffCSP, and InvDesFlow-AL achieve better novelty, but the generated structures often exhibit significantly reduced symmetry, which can lead to increased computational difficulty or instability in subsequent DFT calculations. Overall, there is a trade-off between novel material discovery and computational tractability across different generative architectures, and model selection should be guided by the specific research objectives.

\textcolor{black}{We argue that improving the dynamical stability of generated crystal structures can be fundamentally framed as two complementary paradigms: explicit optimization and implicit guidance. On the one hand, reinforcement learning (RL) enables the direct incorporation of dynamical stability into the objective function by treating the generator as a policy network, where phonon-based stability metrics (e.g., a minimum phonon frequency threshold) serve as reward signals. Given the prohibitive cost of high-fidelity phonon calculations, conventional methods such as Proximal Policy Optimization (PPO)~\cite{PPO2017}, which rely on near-instantaneous feedback, are impractical. Instead, Direct Preference Optimization (DPO)~\cite{DPO2023}, which learns from preference pairs constructed from stable and unstable structures, provides a more tractable alternative, allowing the model to both capture the distribution of stable materials and actively avoid unstable ones. Importantly, the large-scale phonon evaluations in PhononBench naturally generate abundant positive--negative structure pairs, offering a ready-to-use and high-quality preference dataset for DPO training.}

\textcolor{black}{On the other hand, for LLM/Transformer-based generators, prompt engineering offers a low-cost yet effective mechanism for implicit control. Empirically, incorporating prompts such as chemical formulas, space group information, and manually designed instructions leads to substantial improvements in dynamical stability, with the best-performing model, LLaMA2-70B, achieving an $\sim$18\% gain over its no-prompt baseline. This suggests that semantic priors can effectively constrain the generative distribution toward stable regions. Looking forward, physically informed prompts, such as ``stability-first'' or ``low-imaginary-frequency-first,'' may further steer generation toward dynamically stable solutions without additional computational overhead. Taken together, the integration of explicit stability-driven optimization and implicit semantic guidance provides a unified and scalable pathway for generating dynamically stable crystal structures.}

\textcolor{black}{Reliable evaluation of dynamical stability is therefore essential for assessing the practical usefulness of structures generated by these models. In this work, such evaluation relies on large-scale phonon calculations using the MatterSim potential. One possible explanation for the strong phonon prediction performance of MatterSim lies in the distribution of its training data. Compared with many recent universal machine-learning interatomic potentials that are primarily trained on relaxed structures and near-equilibrium configurations, the MatterSim dataset spans a broader thermodynamic space and includes a larger number of perturbed atomic configurations. Such broader coverage of configurational space may help the model better capture the local curvature of the potential energy surface. Because phonon spectra are governed by the second derivatives of the potential energy surface with respect to atomic displacements, training data containing diverse off-equilibrium configurations can improve the reliability of phonon stability predictions. Future work may further explore training strategies that explicitly incorporate off-equilibrium configurations to improve phonon-related predictions.}

\section*{Methods}
\label{sec-method}

\textcolor{black}{To reliably assess the dynamical stability of the crystal structures generated in this work, we computed their phonon spectra using the MatterSim-v1 uMLIP. Phonon spectra depend on the second derivatives of the potential energy surface and therefore require high force accuracy under small atomic displacements. Within the harmonic approximation, the interatomic force constant matrix is defined as
\begin{equation}
\Phi_{ij}^{\alpha\beta} =
\frac{\partial^2 E}{\partial u_i^\alpha \partial u_j^\beta},
\end{equation}
where $E$ is the total energy and $u_i^\alpha$ denotes the displacement of atom $i$ along Cartesian direction $\alpha$. The phonon frequencies are obtained by diagonalizing the dynamical matrix
\begin{equation}
D_{ij}^{\alpha\beta}(\mathbf{q}) =
\frac{1}{\sqrt{m_i m_j}}
\sum_{\mathbf{R}}
\Phi_{ij}^{\alpha\beta}(\mathbf{R})
e^{i\mathbf{q}\cdot\mathbf{R}}.
\end{equation}
Accurate prediction of phonon spectra therefore requires uMLIP to maintain reliable force predictions under small atomic displacements. Recent large-scale phonon benchmarks have shown that MatterSim achieves competitive and well-balanced performance (Table~\ref{tab:phonon_confusion} and Table~\ref{tab:inference_speed}) in predicting dynamical stability, motivating its use in this study.}

A recent large-scale benchmark~\cite{PhonnBench} conducted by Miguel A. L. Marques and co-workers demonstrated that, among seven state-of-the-art uMLIPs, MatterSim-v1 delivers the highest accuracy across key phonon-related properties—including phonon frequencies, phonon DOS, free energy, and heat capacity—with errors even smaller than those arising from the choice of different DFT functionals (e.g., PBE versus PBEsol). Moreover, it achieves an accuracy of approximately 95\% in classifying dynamical stability. 
\textcolor{black}{To further validate the reliability of MatterSim for phonon calculations, we examined its performance on the recent phonon benchmark dataset proposed by Miguel A. L. Marques et al.~\cite{PhonnBench}. The benchmark assesses the ability of uMLIPs to predict dynamical stability via full phonon-spectrum calculations. MatterSim demonstrates balanced performance on this benchmark and ranks among the most accurate models for predicting phonon stability. As shown in Table~\ref{tab:phonon_confusion}, we include the performance of the recently released PET-OAM-XL~\cite{PET-MAD} on the full benchmark dataset, enabling a more comprehensive comparison with state-of-the-art uMLIPs. To enable broader comparison under affordable computational cost, we further evaluate PET-OAM-XL and eSEN-30M-OAM~\cite{esen-30m} on a representative subset of the benchmark. Despite being smaller in model size, MatterSim achieves competitive performance in predicting phonon stability, while maintaining significantly higher computational efficiency. In addition to predictive accuracy, inference efficiency is critical for large-scale phonon calculations. As shown in Table~\ref{tab:inference_speed}, MatterSim exhibits substantially faster inference speed compared with recently released uMLIPs, enabling phonon calculations for more than 100,000 structures in this study.}
Traditional DFT phonon calculations require extensive force-constant evaluations and are therefore prohibitively expensive for high-throughput screening at the scale of tens of thousands of materials. Given the validated accuracy and efficiency of MatterSim-v1, we adopt this model to perform large-scale phonon and dynamical-stability evaluations of the generated structures, enabling reliable and efficient stability screening at unprecedented scale. 

We developed a high-throughput phonon calculation workflow based on Phonopy and the MatterSim-v1 universal machine learning interatomic potential. Crystal structure files (CIF, POSCAR/CONTCAR, etc.) were first converted to \textsc{PhonopyAtoms} objects using a custom batch script, and $2\times2\times2$ supercells were generated to produce compressed \texttt{phonopy.yaml.bz2} input files. Phonon calculations were then performed using MatterSim-v1 for both geometry relaxation and force constant evaluation. Specifically, initial structures were reconstructed from the reference Phonopy files and optimized with the FIRE algorithm while preserving crystal symmetry (force convergence criterion: 0.005~eV/\AA). Displaced supercells (0.01~\AA) were generated by Phonopy, and atomic forces were computed using MatterSim-v1, corrected for translational drift, and used to construct and symmetrize the force constant matrices. High-symmetry paths were automatically generated using \textsc{Seekpath}, and phonon band structures were obtained via Fourier interpolation. \textcolor{black}{Dynamical stability was assessed by checking for imaginary modes. Unless otherwise specified, all analyses in the main text adopt a threshold of $-0.1$~THz, while results under multiple alternative thresholds are provided in the SM.} This high-throughput workflow enables the dynamical stability assessment of tens of thousands of structures, providing an efficient and reliable basis for evaluating phonon properties of generative-model-derived crystals.

\section*{Data availability}
\textcolor{black}{All datasets used in this work are publicly available. The PhononBench datasets, including force constants (YAML format), phonon band structures (NPZ format), and relaxed crystal structures, can be accessed via Zenodo:
\begin{itemize}
    \item \url{https://zenodo.org/records/19328178}
    \item \url{https://zenodo.org/records/19338671}
    \item \url{https://zenodo.org/records/19317118}
    \item \url{https://zenodo.org/records/18185662}
    \item \url{https://zenodo.org/records/19395921} 
\end{itemize}
Additional datasets (Alex20 and MP20) are available at:
\begin{itemize}
    \item \url{https://zenodo.org/records/19346093}
\end{itemize}}

\section*{Code availability} 

The GitHub repository (\url{https://github.com/xqh19970407/PhononBench}) provides code and workflows only. In addition, we have developed a web-based service (\url{http://phononbench.cn/}) for DFT practitioners that enables rapid phonon spectrum calculations, supports user-uploaded arbitrary crystal structures, and provides downloadable phonon results, figures, and relaxed structures. The service has been released via the same GitHub repository.

\bibliographystyle{unsrt}
\bibliography{references}

\vspace{36pt}
\noindent\textbf{Acknowledgement:}
The work is supported by Beijing Natural Science Foundation(No.Z250005), the National Natural Science Foundation of China (No.62476278, No.11934020), and the National Key R\&D Program of China (Grants No. 2024YFA1408601). Computational resources have been provided by the Physical Laboratory of High Performance Computing at Renmin University of China. \textcolor{black}{We gratefully acknowledge Zhenfeng Ouyang for insightful discussions on the physical interpretation of the results, and Jing Jiang for valuable discussions on DFT calculations.}
\\

\noindent\textbf{Corresponding authors:} Correspondence and requests for materials should be addressed to Ze-Feng Gao (zfgao@ruc.edu.cn) and Zhong-Yi Lu (zlu@ruc.edu.cn). \\

\noindent\textbf{Competing interests:}
The authors declare no competing interests.\\

\noindent\textbf{Supplementary materials:}
The supplementary materials is attached.
\clearpage
\appendix

\section*{Appendix Contents}

\begin{enumerate}
    \item \hyperref[app:benchmark]{Benchmarking the Accuracy and Efficiency of Phonon Calculations Using uMLIPs}
    \begin{enumerate}
        \item {Phonon Prediction on 9958 MP Materials Using SOTA Models}
        \item {Inference Efficiency}
        \item {DFT Validation on 120 AI-Generated Materials}
    \end{enumerate}

    \item \hyperref[app:physical]{Physical Origins of Dynamical Instability with Multi-Threshold Analysis}
    \begin{enumerate}
        \item Element
        \item Chemical Complexity
        \item Coordination Environments
        \item Formation Energy
        \item Metal vs. Insulator
        \item Space Group Symmetry
    \end{enumerate}

    \item \hyperref[app:scale]{Effect of Training Data Scale on Dynamical Stability}

    \item \hyperref[app:llm]{Enhancing LLM-Based Crystal Generation through Prompting and Scaling}
    \begin{enumerate}
        \item {CrystaLLM-large}
        \item {LLaMA-2 70B}
    \end{enumerate}

    \item \hyperref[app:summary]{Summary of Large-Scale Crystal Generation and Dynamical Stability}

    \item \hyperref[app:convergence]{Convergence Analysis of Dynamical Stability Rate}

    \item \hyperref[app:generation]{Summary of Crystal Generation}

    \item \hyperref[app:bandgap]{Crystal Relaxation and Dynamical Stability Statistics Across Bandgap Ranges}

    \item \hyperref[app:data]{Data Download}
\end{enumerate}

\newpage
\section{\textcolor{black}{Benchmarking the Accuracy and Efficiency of Phonon Calculations Using uMLIPs}}
\label{app:benchmark}
\subsection{\textcolor{black}{Phonon Prediction on 9958 MP Materials Using SOTA Models}}
\begin{figure}[H]
		\centering  
		\includegraphics[width=1.0\linewidth]{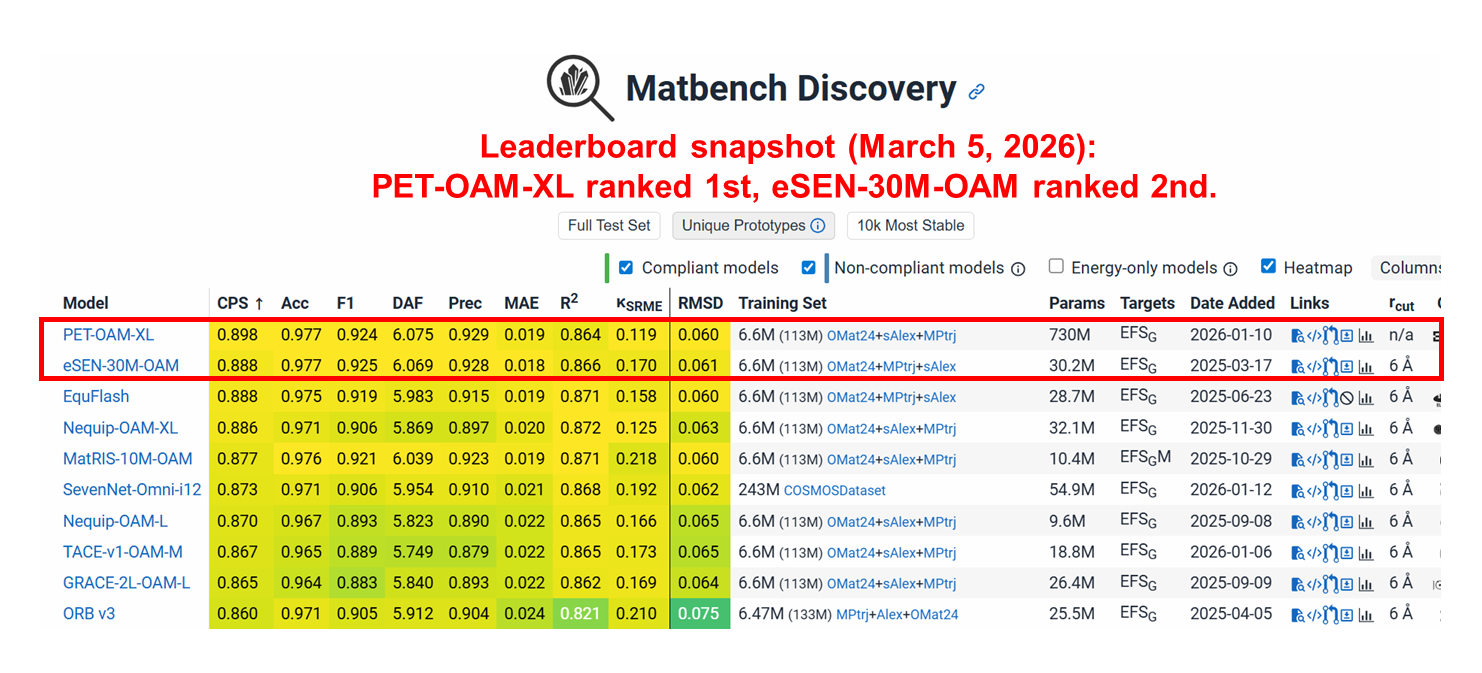}
        
		\caption{Matbench Discovery Leaderboard (March 5, 2026).}
		\label{mat-bench} 
\end{figure}

 \textcolor{black}{First, it is important to note that the Matbench-Discovery leaderboard does not directly reflect the performance of universal machine learning interatomic potentials (uMLIPs) in phonon prediction tasks, because it primarily evaluates models based on energy and force prediction accuracy on near-equilibrium structures. In contrast, phonon spectra depend on the second derivatives of the potential energy surface and therefore require the model to maintain high force accuracy under small atomic displacements. To further clarify this point, as shown in Figure~\ref{mat-bench}, we have additionally evaluated two recently released models that currently rank first and second on the Matbench Discovery leaderboard (snapshot taken on March 5, 2026): PET-OAM-XL~\cite{pet-oam} (released on 2026-01-10) and eSEN-30M-OAM~\cite{eSEN-30M-OAM} (released on 2025-03-17). We included these models in additional tests specifically related to phonon prediction.}

\begin{figure}[H]
		\centering  
		\includegraphics[width=1.0\linewidth]{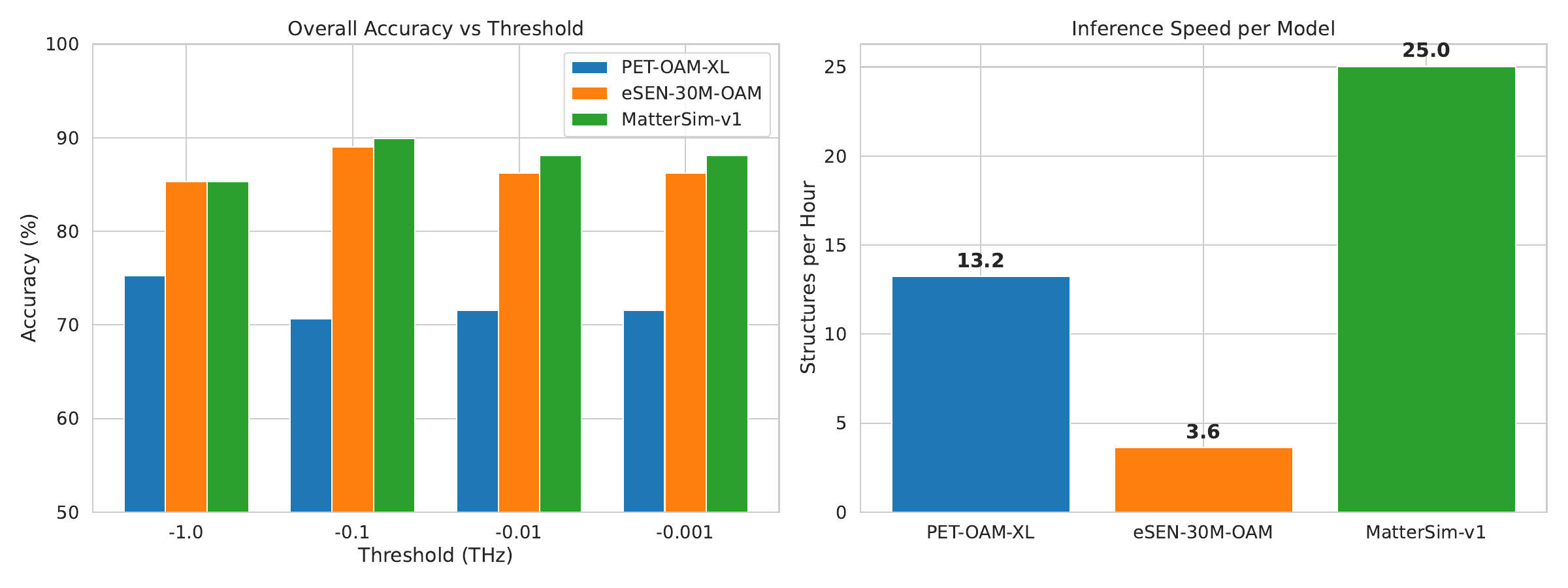}
        
		\caption{\textcolor{black}{Comparison of dynamical stability prediction for 120 AI-generated materials not present in existing databases, using DFT phonon calculations and three interatomic potential models. Left: Overall accuracy (\%) of PET-OAM-XL, eSEN-30M-OAM, and MatterSim-v1 at different phonon frequency thresholds, quantifying agreement with DFT-based phonon stability. Right: Inference speed of each model, measured as the number of structures evaluated per hour. MatterSim-v1 provides the fastest predictions, while eSEN-30M-OAM is the slowest, highlighting the trade-off between prediction accuracy and computational efficiency for dynamical stability assessments.}}
		\label{accuracy-speed} 
\end{figure}

\textcolor{black}{As shown in Table~\ref{tab:phonon_confusion}, the upper part of the table reports the phonon prediction benchmark results from Ref. 39, which we adopt as the baseline for this work. In this benchmark, phonon spectra calculated using the PBE functional are taken as the reference. The benchmark dataset contains 8,189 stable structures and 1,769 unstable structures, where the stability labels are determined from the corresponding DFT phonon spectra. The predictions of different models are evaluated against these DFT-derived labels, and the corresponding confusion matrices are constructed. From these benchmark results, MatterSim exhibits the best performance in identifying stable structures (TS) while maintaining a reasonable ability to detect unstable structures (TU), indicating competitive and well-balanced performance for predicting phonon stability.}

\begin{table}[H]
\centering
\color{black}
\caption{\textcolor{black}{Normalized confusion matrix (\%) for predicted dynamical stability.
TS: true stable, FU: false unstable, TU: true unstable, FS: false stable.
The Benchmark of Ref. 39 contains 8189 stable and 1769 unstable structures.
The subset contains 50 stable and 50 unstable structures.$\clubsuit$ denotes additional tests introduced in this work to address the reviewer’s concern.}}
\begin{tabular}{llcccc}
\toprule
Dataset & Model & TS$\uparrow$ (\%) & FU$\downarrow$ (\%) & TU$\uparrow$ (\%) & FS$\downarrow$ (\%) \\
\midrule
\multirow{9}{*}{Ref. 39}
 & M3GNet~\cite{m3gnet} & 87 & 13 & 73 & 27 \\
 & CHGNet~\cite{Deng2023CHGNet} & 77 & 23 & 73 & 27 \\
 & MACE-MP-0~\cite{MACE-mp-0} & 95 & 5 & 73 & 27 \\
 & SevenNet-0~\cite{Park2024-SevenNet-0} & 81 & 19 & 80 & 20 \\
 & ORB~\cite{orb-2024} & 15 & 85 & 92 & 8 \\
 & eqV2-M~\cite{eqv2-omat24} & 7 & 93 & \textbf{94} & \textbf{6} \\
 & PBESol (DFT reference) & 97 & 3 &  76 & 24 \\
 
 & MatterSim-v1~\cite{mattersim} & \textbf{95} & \textbf{5} & 75 & 25 \\
 &  PET-OAM-XL~\cite{pet-oam} $\clubsuit$& 63 & 37 & 91 & 9 \\
\midrule
\multirow{4}{*}{Subset (100)}
 & eSEN-30M-OAM~\cite{eSEN-30M-OAM}$\clubsuit$ & 76 & 24 & 84 & 16 \\
  & PET-OAM-XL~\cite{pet-oam}$\clubsuit$ & 68 & 32 & \textbf{98} & \textbf{2} \\
  & PBESol (DFT reference) $\clubsuit$& 100 & 0 &  80 & 20 \\
 & MatterSim-v1~\cite{mattersim} $\clubsuit$& \textbf{96} & \textbf{4} & 80 & 20 \\
\bottomrule
\end{tabular}
\label{tab:phonon_confusion}
\end{table}

\textcolor{black}{In addition to the baseline results reported in Ref.~39, we further include the performance of the recently released PET-OAM-XL model on the full benchmark dataset, which currently ranks first on the Matbench-Discovery leaderboard ($\clubsuit$ denotes additional tests introduced in this work to address the reviewer’s concern.). As shown in the last row of the upper part of Table~\ref{tab:phonon_confusion}, PET-OAM-XL achieves 63\% TS and 91\% TU, indicating a strong capability in identifying dynamically unstable structures, but a comparatively lower accuracy in recognizing stable ones. This imbalance suggests that, despite its leading performance in energy and force prediction benchmarks, its performance in phonon-based stability classification is less well balanced than some existing models such as MatterSim. To further investigate this behavior and to enable a broader comparison with newly released models under affordable computational cost, we additionally construct a representative subset for systematic evaluation, as discussed below.}

\textcolor{black}{The lower part of the Table~\ref{tab:phonon_confusion} presents the additional results introduced in this work. Specifically, we randomly selected 50 stable structures and 50 unstable structures from the above benchmark dataset to form a subset of 100 structures, which was used to evaluate the phonon prediction performance of newly released models. As shown by the results, the performance of MatterSim on this subset is highly consistent with its performance on the full benchmark dataset. For example, MatterSim achieves 95\% TS and 5\% FU on the full benchmark dataset, compared with 96\% TS and 4\% FU on the subset. This consistency suggests that the constructed subset reasonably preserves the statistical characteristics of the original dataset, and therefore provides a representative and reliable basis for comparing newly released models. Given the high computational cost of phonon calculations, evaluating newly released models on a representative subset provides a practical yet informative comparison.
Within this subset, both PET-OAM-XL and eSEN-30M-OAM exhibit noticeably weaker performance than MatterSim in identifying stable structures. Specifically, MatterSim achieves a TS of 96\%, whereas eSEN-30M-OAM and PET-OAM-XL obtain only 76\% and 68\%, respectively. This implies that PET-OAM-XL incorrectly classifies approximately 32\% of truly stable structures as unstable (FU), which is substantially higher than the 4\% observed for MatterSim. Such errors are particularly critical in high-throughput materials screening, because falsely predicting stable materials as unstable may prematurely eliminate potentially viable candidates, thereby significantly reducing the efficiency of the screening process. On the other hand, PET-OAM-XL achieves a high TU value of 98\%, indicating strong capability in detecting unstable structures; however, its weaker performance in recognizing stable structures suggests that the model tends to classify structures as unstable. In contrast, MatterSim maintains a high stable-structure identification rate (TS = 96\%) while also preserving a reasonable ability to detect unstable structures (TU = 80\%), achieving a more balanced predictive performance between the two tasks. Therefore, from the perspective of the overall confusion matrix, MatterSim demonstrates more stable and reliable performance in distinguishing between stable and unstable structures.}

\textcolor{black}{We compare the predictions of MatterSim with DFT results and analyze the error relative to different exchange--correlation functionals. As shown in Table~\ref{tab:phonon_confusion}, the discrepancy between MatterSim and DFT (using the PBE functional) is comparable to the intrinsic variation between different DFT functionals, such as PBE and PBESol, both on the full benchmark dataset and on the representative subset constructed in this work. This observation indicates that the prediction error introduced by MatterSim lies within the uncertainty range of first-principles methods themselves. In other words, the deviation between MatterSim and DFT is on the same order as the variation arising from the choice of exchange--correlation functional.}

\subsection{\textcolor{black}{Inference Efficiency}}
\textcolor{black}{In addition, as shown in Table~\ref{tab:inference_speed}, MatterSim-v1 also exhibits clear advantages in inference efficiency. For the same set of 100 structures, MatterSim completes all calculations in only 3.99 hours, whereas PET-OAM-XL and eSEN-30M-OAM require 7.55 hours and 27.53 hours, respectively. Consequently, the inference speed of MatterSim is approximately twice that of PET-OAM-XL and about seven times faster than that of eSEN-30M-OAM. The latter employs a higher-order equivariant graph neural network architecture, which introduces a large number of tensor product operations. This efficiency advantage is particularly important for high-throughput phonon stability screening, where a large number of candidate structures typically need to be evaluated in practical materials discovery workflows.}

\begin{table}[h]
\centering
\color{black}
\caption{\textcolor{black}{Inference efficiency comparison for phonon calculations on 100 structures.}}
\begin{tabular}{lccc}
\hline
Model & Total Time (100 structures) & Avg Time / Structure  \\
\hline
PET-OAM-XL & 7 h 33 m 13 s & 271.9 s \\
eSEN-30M-OAM & 27 h 31 m 59 s & 990.0 s \\
MatterSim-v1 & 3 h 59 m 36 s & \textbf{143.8 s}  \\
\hline
\end{tabular}
\label{tab:inference_speed}
\end{table}

\textcolor{black}{In addition, differences in the training data distributions may also contribute to the observed performance. Many recent uMLIPs are trained using combinations of the OMat24, sAlex, and MPtrj datasets, which mainly contain relaxed structures, rattled configurations, and short molecular dynamics trajectories near equilibrium. In contrast, the MatterSim dataset contains approximately 17 million structures and spans a much wider thermodynamic space, covering temperatures from 0–5000 K and pressures up to 1000 GPa. This broader coverage of configurational space introduces a large number of perturbed configurations, which can help the model better learn the local curvature of the potential energy surface. Since phonon spectra are determined by the second derivatives of the potential energy surface with respect to atomic displacements, such training data characteristics are beneficial for improving the reliability of phonon stability predictions. These results collectively indicate that, despite the emergence of newer uMLIPs, MatterSim remains a reliable and efficient model for phonon stability prediction and is therefore an appropriate choice for the present benchmark study.}

\textcolor{black}{Therefore, the choice of MatterSim-v1 in this work is motivated by three main considerations:
(i) its strong and balanced performance in phonon stability prediction,
(ii) its significantly higher inference efficiency for large-scale screening,
and (iii) its training data distribution that better captures the local curvature of the potential energy surface relevant to phonon calculations.
Taken together, these factors make MatterSim-v1 a practical and reliable backbone for the large-scale phonon benchmarking conducted in this study. The corresponding discussions have been added to the revised manuscript.}

\subsection{\textcolor{black}{DFT Validation on 120 AI-Generated Materials}}
\textcolor{black}{Beyond existing materials, we constructed an independent validation dataset by randomly sampling 120 AI-generated structures (60 stable and 60 unstable) to assess the reliability of MatterSim. All materials were subjected to full DFT-based phonon calculations. The resulting dataset spans a broad chemical space, including light-element systems (e.g., Li, B, H), transition-metal compounds (e.g., Fe, Co, Ni, Pd), and heavy-element-containing materials (e.g., Au, Pb, and rare-earth elements). As shown in Table~\ref{dfpt-tab-120-sub}, representative results are presented, while the complete dataset has been made publicly available. The selected materials span compositions from binary to quaternary systems and exhibit diverse bonding characteristics, including metallic, ionic, and covalent interactions, while also covering a wide range of space groups. This diversity ensures an unbiased evaluation and enables a comprehensive assessment of model performance across different chemical and structural spaces. In the table, a checkmark ($\checkmark$) indicates agreement between MatterSim and DFT under the given imaginary frequency threshold, while a cross ($\times$) indicates disagreement. Overall, MatterSim demonstrates strong capability in classifying dynamical stability across different threshold settings. For completeness, a small proportion of misclassified materials is also included at the end of the table, though their overall fraction remains low.}

\begin{table}[H]
\centering
\color{black}
\caption{\textcolor{black}{Comparison of minimum phonon frequencies and dynamical stability predictions between MatterSim and DFT for representative materials. A checkmark ($\checkmark$) indicates agreement between MatterSim and DFT under the given imaginary frequency threshold, while a cross ($\times$) indicates disagreement.}}
\label{tab:ai_dft_cases}
\begin{tabular}{lccccccc}
\toprule
Formula & Space Group & No. & $\omega_{\min}^{\text{DFT}}$ & $\omega_{\min}^{\text{AI}}$ & $-0.001$ & $-0.1$ & $-1.0$ \\
\midrule
Ac$_3$LaN$_4$ & Pm$\bar{3}$m & 221 & -8.49 & -5.88 & $\checkmark$ & $\checkmark$ & $\checkmark$ \\
AcNd$_3$Co & Pm$\bar{3}$m & 221 & -2.74 & -2.32 & $\checkmark$ & $\checkmark$ & $\checkmark$ \\
Al$_3$CdS$_4$ & P$\bar{4}$3m & 215 & -5.16 & -2.48 & $\checkmark$ & $\checkmark$ & $\checkmark$ \\
AlF$_2$ & Pa$\bar{3}$ & 205 & -12.69 & -13.45 & $\checkmark$ & $\checkmark$ & $\checkmark$ \\
Al$_3$W & Pm$\bar{3}$n & 223 & -3.94 & -3.94 & $\checkmark$ & $\checkmark$ & $\checkmark$ \\
Be$_3$Cu$_4$ & P$\bar{4}$3m & 215 & -6.90 & -8.62 & $\checkmark$ & $\checkmark$ & $\checkmark$ \\
BeS$_2$ & Pa$\bar{3}$ & 205 & 0.00 & 0.00 & $\checkmark$ & $\checkmark$ & $\checkmark$ \\
BeCo$_3$H & Pm$\bar{3}$m & 221 & -3.41 & -2.09 & $\checkmark$ & $\checkmark$ & $\checkmark$ \\
CaF$_2$ & P4$_2$/mnm & 136 & 0.00 & 0.00 & $\checkmark$ & $\checkmark$ & $\checkmark$ \\
Ca$_3$EuF & Pm$\bar{3}$m & 221 & -2.41 & -4.02 & $\checkmark$ & $\checkmark$ & $\checkmark$ \\
CaAc$_3$ & Pm$\bar{3}$m & 221 & 0.00 & 0.00 & $\checkmark$ & $\checkmark$ & $\checkmark$ \\
CaFe$_2$ & P6/mmm & 191 & 0.00 & 0.00 & $\checkmark$ & $\checkmark$ & $\checkmark$ \\
Co$_5$Ni & P$\bar{6}$2m & 189 & 0.00 & 0.00 & $\checkmark$ & $\checkmark$ & $\checkmark$ \\
Cr$_3$PdS$_4$ & P$\bar{4}$3m & 215 & -2.78 & -1.66 & $\checkmark$ & $\checkmark$ & $\checkmark$ \\
CsGa$_3$Hg & Pm$\bar{3}$m & 221 & -1.96 & -1.74 & $\checkmark$ & $\checkmark$ & $\checkmark$ \\
FeCu$_3$Se$_4$ & P$\bar{4}$3m & 215 & 0.00 & 0.00 & $\checkmark$ & $\checkmark$ & $\checkmark$ \\
GaH$_4$Se & P$\bar{6}$m2 & 187 & -193.22 & -5.76 & $\checkmark$ & $\checkmark$ & $\checkmark$ \\
GdTmGa$_6$ & Pm$\bar{3}$ & 200 & -2.38 & -1.58 & $\checkmark$ & $\checkmark$ & $\checkmark$ \\
Hf$_3$Pt & Pm$\bar{3}$n & 223 & 0.00 & 0.00 & $\checkmark$ & $\checkmark$ & $\checkmark$ \\
K$_3$V & P6$_3$/mmc & 194 & 0.00 & 0.00 & $\checkmark$ & $\checkmark$ & $\checkmark$ \\
LiAl$_3$ & P6$_3$/mmc & 194 & 0.00 & 0.00 & $\checkmark$ & $\checkmark$ & $\checkmark$ \\
LiBH$_3$ & P6$_3$22 & 182 & -35.37 & -16.51 & $\checkmark$ & $\checkmark$ & $\checkmark$ \\
LiF$_4$ & P4/mmm & 123 & -5.35 & -3.38 & $\checkmark$ & $\checkmark$ & $\checkmark$ \\
Li$_2$Mg$_2$Al & Pmm2 & 25 & -2.91 & -9.59 & $\checkmark$ & $\checkmark$ & $\checkmark$ \\
LiMg$_2$ & P6$_3$/mmc & 194 & -2.01 & -1.81 & $\checkmark$ & $\checkmark$ & $\checkmark$ \\
Li$_3$CdCuF$_4$ & P$\bar{4}$3m & 215 & -4.96 & -4.55 & $\checkmark$ & $\checkmark$ & $\checkmark$ \\
Li$_3$Mg & Pm$\bar{3}$m & 221 & 0.00 & 0.00 & $\checkmark$ & $\checkmark$ & $\checkmark$ \\
LiIn & P$\bar{4}$3m & 215 & -1.92 & -2.08 & $\checkmark$ & $\checkmark$ & $\checkmark$ \\
Li$_3$Au & Pm$\bar{3}$n & 223 & 0.00 & 0.00 & $\checkmark$ & $\checkmark$ & $\checkmark$ \\
Li$_3$Pb & Pm$\bar{3}$n & 223 & 0.00 & 0.00 & $\checkmark$ & $\checkmark$ & $\checkmark$ \\
KSeI$_3$ & Pm$\bar{3}$m & 221 & -0.69 & -1.64 & $\checkmark$ & $\checkmark$ & $\times$ \\
KVAs & P$\bar{6}$m2 & 187 & -0.77 & 0.00 & $\times$ & $\times$ & $\checkmark$ \\
BaMg$_3$Ir & Pm$\bar{3}$m & 221 & -0.24 & 0.00 & $\times$ & $\times$ & $\checkmark$ \\
Li$_2$ScPd$_3$ & P6mm & 183 & -0.05 & -1.89 & $\checkmark$ & $\times$ & $\times$ \\
Ir$_3$Se$_4$ & P$\bar{4}$3m & 215 & -3.70 & 0.00 & $\times$ & $\times$ & $\times$ \\
Fe$_3$Cu$_2$ & P4/m & 83 & -3.11 & 0.00 & $\times$ & $\times$ & $\times$ \\
BeF$_4$ & P$\bar{4}$3m & 215 & 0.00 & -3.54 & $\times$ & $\times$ & $\times$ \\
Hf$_6$NpSn & Pm$\bar{3}$ & 200 & -1.59 & 0.00 & $\times$ & $\times$ & $\times$ \\
Ac$_3$Tb & Pm$\bar{3}$m & 221 & -1.72 & 0.00 & $\times$ & $\times$ & $\times$ \\
\bottomrule
\label{dfpt-tab-120-sub}
\end{tabular}
\end{table}

\textcolor{black}{To further quantify the overall statistical trends, we additionally construct a confusion matrix, as summarized in Table~\ref{tab:ai_dft_confusion}. Notably, the datasets in Table~\ref{tab:ai_dft_confusion} and Table~\ref{tab:phonon_confusion} differ fundamentally in nature. The former consists of 120 AI-generated materials validated by DFT, which explicitly fall into the ``not seen during training'' regime and therefore represent a stringent out-of-distribution (OOD) test. On the OOD dataset, MatterSim maintains consistently strong performance. For example, under a strict threshold ($-0.001$ THz), the true stable rate (TS) reaches 97.62\% with a low false unstable rate (FU) of 2.38\%. Under a more relaxed threshold ($-1.0$ THz), the true unstable rate (TU) further improves to 79.66\%, indicating robust capability in identifying unstable structures. Importantly, comparing these results with those obtained on in-distribution datasets (Table~\ref{tab:phonon_confusion}) shows that MatterSim achieves consistent and stable performance across both regimes. This suggests that the model captures physically meaningful structure–property relationships rather than merely relying on correlations within the training distribution. Therefore, DFT-based validation demonstrates that MatterSim maintains strong generalization ability and predictive reliability on out-of-distribution structures, mitigating concerns about potential bias propagation.}

\begin{table}[H]
\centering
\color{black}
\caption{\textcolor{black}{Normalized confusion matrices (\%) comparing dynamical stability predictions from MatterSim against DFT reference calculations under different imaginary frequency thresholds. A structure is considered stable if its minimum phonon frequency exceeds the specified threshold. All results are evaluated on 120 DFT-validated materials.}}
\label{tab:ai_dft_confusion}
\begin{tabular}{lcccc}
\toprule
Threshold (THz) & TS$\uparrow$ (\%) & FU$\downarrow$ (\%) & TU$\uparrow$ (\%) & FS$\downarrow$ (\%) \\
\midrule
$-0.001$ & \textbf{97.62} & \textbf{2.38} & 74.63 & 25.37 \\
$-0.010$ & \textbf{97.62} & \textbf{2.38} & 74.63 & 25.37 \\
$-0.100$ & 95.35 & 4.65 & 74.24 & 25.76 \\
$-0.500$ & 95.45 & 4.55 & 75.38 & 24.62 \\
$-1.000$ & 92.00 & 8.00 & \textbf{79.66} & \textbf{20.34} \\
\bottomrule
\label{cm-table-120}
\end{tabular}
\end{table}

\begin{figure}[H]
		\centering  
		\includegraphics[width=1.0\linewidth]{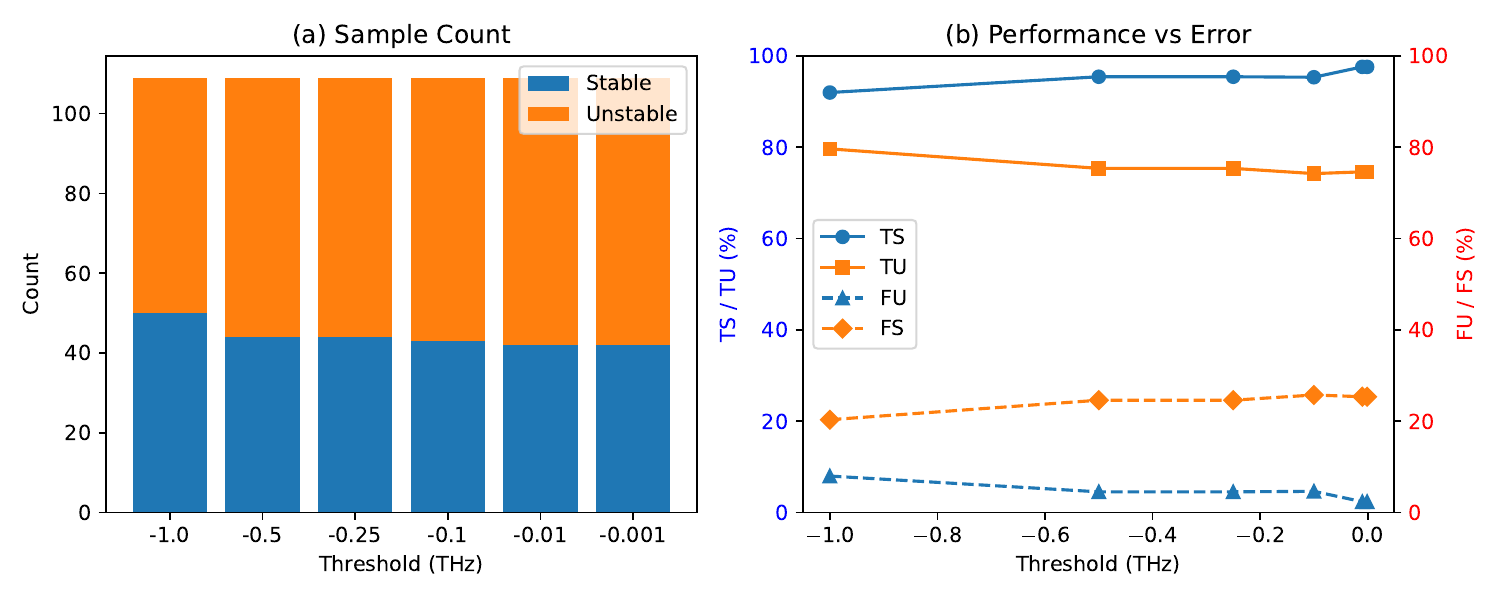}
        
\caption{
\textcolor{black}{Visualization of dynamical stability predictions under different imaginary frequency thresholds. 
(a) Sample count number of stable and unstable structures at each threshold. 
(b) Performance metrics: true stable (TS) and true unstable (TU) rates are plotted on the left y-axis (blue), while false unstable (FU) and false stable (FS) rates are plotted on the right y-axis (red). 
This dual-axis representation highlights both the distribution of samples and the model's predictive performance across thresholds.
All results are based on DFT-validated AI-generated materials.}
}
\label{fig:dft-stability-performance}
\end{figure}

\textcolor{black}{Figure~\ref{fig:dft-stability-performance}(b) further evaluates the consistency between MatterSim predictions and DFT results across different imaginary frequency thresholds. Specifically, the true stable (TS) and true unstable (TU) rates are shown on the left axis, while the false unstable (FU) and false stable (FS) rates are shown on the right axis. Notably, the overall prediction performance remains highly stable across a wide range of thresholds. For example, the variation in accuracy between commonly used thresholds (e.g., -0.1 THz and -1.0 THz) is within a few percent, indicating that the model’s predictions are largely insensitive to the specific threshold choice. This robustness suggests that the agreement between MatterSim and DFT does not arise from threshold-specific tuning, but instead reflects consistent and physically meaningful predictions. Combined with the DFT-based validation, these results provide strong evidence that MatterSim achieves reliable generalization on out-of-distribution AI-generated structures.}

\textcolor{black}{We perform a comprehensive multi-model comparison across three representative phonon prediction models, as summarized in Table~\ref{dft-120-mul-model}. This analysis enables us to explicitly identify model-dependent biases and assess whether MatterSim exhibits any systematic deviation. First, the results reveal clear and distinct bias patterns across different models.
PET-OAM-XL consistently achieves very high true unstable (TU) rates (e.g., 97.01\% at $-0.001$ THz), but at the cost of extremely low true stable (TS) rates (30.95\%). This indicates a strong bias toward predicting structures as unstable, leading to a large number of false unstable (FU) predictions. In other words, PET-OAM-XL tends to overestimate instability, which may significantly underestimate the pool of viable stable candidates. In contrast, eSEN-30M-OAM achieves a more balanced performance, with relatively high TS (83.33\%) and TU (88.06\%) rates under strict thresholds. However, it still shows a noticeable tendency to classify stable structures as unstable, as reflected by its FU rate (16.67\%), which is substantially higher than that of MatterSim. This suggests a moderate conservative bias toward instability, albeit less severe than PET-OAM-XL.}

\begin{table}[H]
\centering
\color{black}
\small
\caption{\textcolor{black}{Comparison of dynamical stability prediction under different frequency thresholds. 
TS: true stable, FU: false unstable, TU: true unstable, FS: false stable. 
BA: balanced accuracy defined as $(\mathrm{TS}+\mathrm{TU})/2$. 
Acc: overall accuracy. Arrows indicate whether higher ($\uparrow$) or lower ($\downarrow$) values are better.}}
\begin{tabular}{c|c|cccccc}
\hline
Threshold (THz) & Method 
& TS (\%) $\uparrow$ 
& FU (\%) $\downarrow$ 
& TU (\%) $\uparrow$ 
& FS (\%) $\downarrow$ 
& BA (\%) $\uparrow$ 
& Acc (\%) $\uparrow$ \\
\hline

\multirow{3}{*}{-0.001}
&PET-OAM-XL   
& 30.95 & 69.05 & \textbf{97.01} & \textbf{2.99} & 63.98 & 71.56 \\
& eSEN-30M-OAM 
& 83.33 & 16.67 & 88.06 & 11.94 & 85.70 & 86.24 \\
& MatterSim-v1    
& \textbf{97.62} & \textbf{2.38} & 74.63 & 25.37 & \textbf{86.13} & \textbf{88.07} \\

\hline

\multirow{3}{*}{-0.01}
&PET-OAM-XL   
& 30.95 & 69.05 & \textbf{97.01} & \textbf{2.99} & 63.98 & 71.56 \\
& eSEN-30M-OAM 
& 83.33 & 16.67 & 88.06 & 11.94 & 85.70 & 86.24 \\
& MatterSim-v1    
& \textbf{97.62} &\textbf{2.38} & 74.63 & 25.37 & \textbf{86.13} & \textbf{88.07} \\

\hline

\multirow{3}{*}{-0.1}
&PET-OAM-XL   
& 30.23 & 69.77 & \textbf{96.97} & \textbf{3.03} & 63.60 & 70.64 \\
& eSEN-30M-OAM 
& 88.37 & 11.63 & 89.39 & 10.61 & \textbf{88.88} & 88.99 \\
& MatterSim-v1    
& \textbf{95.35} & \textbf{4.65} & 74.24 & 25.76 & 84.80 & \textbf{89.91} \\

\hline

\multirow{3}{*}{-1.0}
&PET-OAM-XL   
& 52.00 & 48.00 & \textbf{94.92} & \textbf{5.08} & 73.46 & 75.23 \\
& eSEN-30M-OAM 
& 90.00 & 10.00 & 81.36 & 18.64 & 85.68 & 85.32 \\
& MatterSim-v1    
& \textbf{92.00} & \textbf{8.00} & 79.66 & 20.34 & \textbf{85.83} & \textbf{85.32} \\

\hline
\end{tabular}
\label{dft-120-mul-model}
\end{table}

\textcolor{black}{MatterSim-v1 exhibits a more balanced behavior. It achieves the highest TS rate across all thresholds (e.g., 97.62\% at $-0.001$ THz), while maintaining competitive TU performance (74.63\%). Importantly, it consistently delivers the best or near-best overall accuracy (Acc) and balanced accuracy (BA), indicating that it does not overly favor either stable or unstable classifications. This suggests that MatterSim does not suffer from the strong instability bias observed in other models, but instead captures a more physically balanced decision boundary. Furthermore, this conclusion is reinforced by the robustness across different thresholds. As the imaginary frequency threshold is relaxed from $-0.001$ to $-1.0$ THz, all models exhibit relatively stable performance trends, but the relative ranking remains consistent: MatterSim maintains top-tier accuracy, while PET-OAM-XL remains strongly biased toward instability. This consistency indicates that the observed differences are not artifacts of a particular threshold choice, but reflect intrinsic model behaviors. In addition to accuracy, we also compare inference efficiency (Figure~\ref{accuracy-speed}). MatterSim achieves the fastest inference speed (approximately 25 structures per hour), while eSEN-30M-OAM—despite comparable accuracy—requires significantly higher computational cost (approximately eight times slower). This highlights an important practical advantage of MatterSim, as it provides both high accuracy and computational efficiency.}

\textcolor{black}{Overall, the multi-model comparison demonstrates that:
(i) systematic biases indeed exist across different models,
(ii) these biases are model-dependent rather than universal, and
(iii) MatterSim does not exhibit the pathological bias toward instability seen in other models.}

\textcolor{black}{Therefore, the agreement between MatterSim and DFT cannot be attributed to a self-reinforcing “echo chamber,” but instead reflects its ability to capture physically meaningful stability trends. This multi-model analysis provides an additional layer of validation and strengthens the credibility of the benchmark. }


\textcolor{black}{As shown in Table~\ref{dfpt-para}, the phonon properties were calculated using density functional perturbation theory (DFPT) as implemented in the Vienna Ab initio Simulation Package (VASP). Force constants were obtained using the linear-response mode (\texttt{IBRION = 8}), and phonon dispersion relations were subsequently evaluated with the PHONOPY package. A $2 \times 2 \times 2$ supercell was used to construct the dynamical matrix. The plane-wave kinetic energy cutoff was set to 700~eV, and a $\Gamma$-centered Monkhorst--Pack $k$-point mesh with a density of approximately $50~\text{\AA}$ in reciprocal space was employed. The electronic self-consistent convergence criterion was $10^{-8}$~eV. Dynamical matrices were then Fourier-interpolated to generate phonon dispersion relations along high-symmetry paths in the Brillouin zone.}
\begin{table}[H]
\small
\centering
\color{black}
\caption{\textcolor{black}{Summary of computational settings for phonon calculations.}}
\label{tab:phonon_settings}
\begin{tabular}{l l}
\toprule
\textbf{Parameter} & \textbf{Setting} \\
\midrule
Method & Density Functional Perturbation Theory (DFPT) \\
Software & VASP (Vienna Ab initio Simulation Package) \\
Force constant calculation & Linear-response mode (\texttt{IBRION = 8}) \\
Phonon analysis & PHONOPY package \\
Supercell size & $2 \times 2 \times 2$ \\
Plane-wave cutoff energy & 700~eV \\
$k$-point mesh & $\Gamma$-centered Monkhorst--Pack grid, $\sim 50~\text{\AA}$ spacing in reciprocal space \\
Electronic convergence criterion & $10^{-8}$~eV \\
Dynamical matrix construction & Fourier interpolation of DFPT-calculated force constants \\
Phonon dispersion & Along high-symmetry paths in the Brillouin zone \\
\bottomrule
\end{tabular}
\label{dfpt-para}
\end{table}

\section{\textcolor{black}{Physical Origins of Dynamical Instability with Multi-Threshold Analysis}}
\label{app:physical}
\subsection{\textcolor{black}{Element}}

\begin{figure}[H]
		\centering  
		\includegraphics[width=1.0\linewidth]{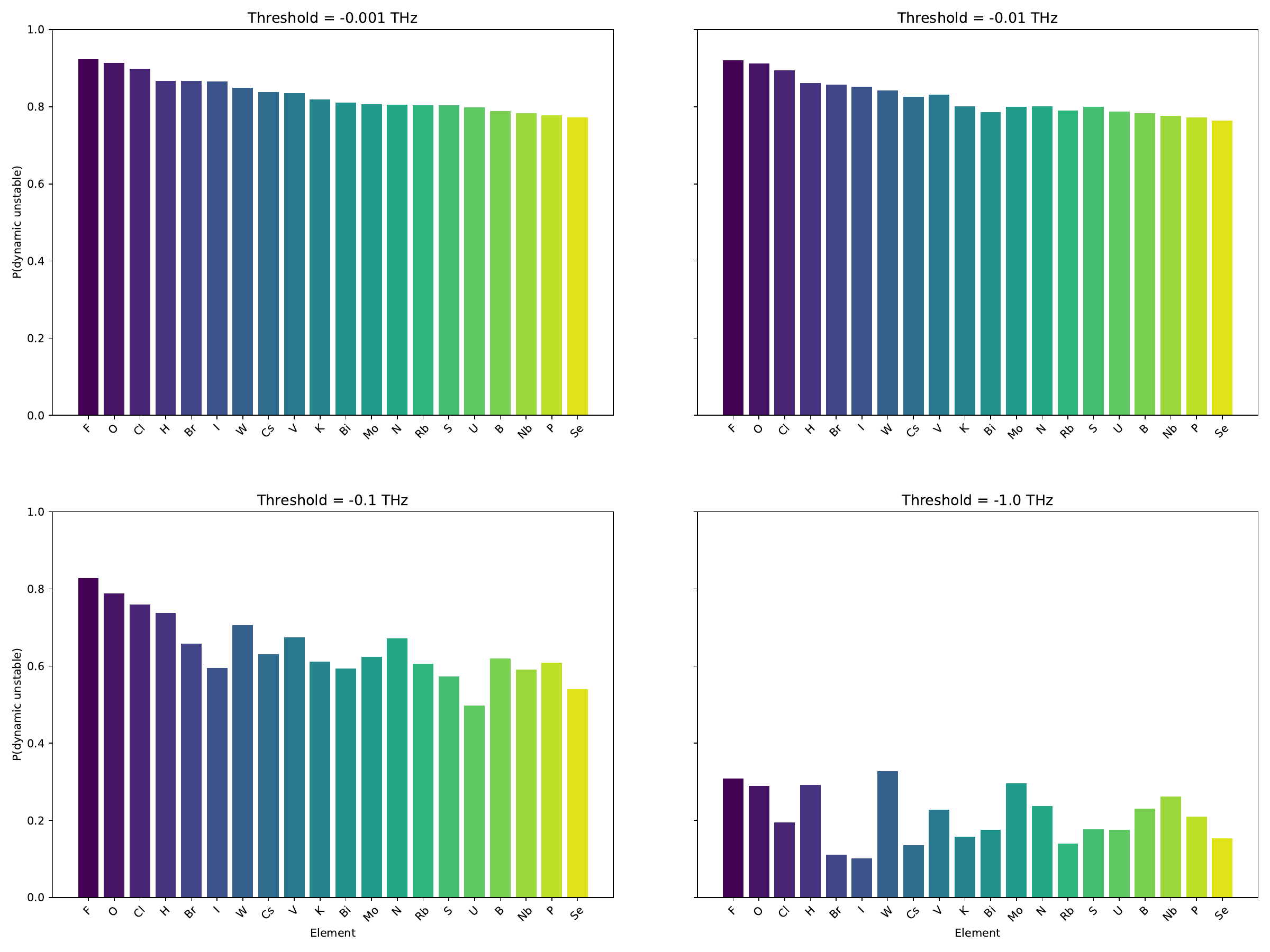}
        \color{black}
\caption{\textcolor{black}{Element-resolved dynamical instability for AI-generated structures across imaginary frequency thresholds ($-0.001$, $-0.01$, $-0.1$, $-1.0$~THz). Elements are ranked by instability probability at the strictest threshold ($-0.001$~THz), with consistent ordering across thresholds. Highly electronegative nonmetals (O, F, Cl) show the highest instability due to directional bonding, alkali metals (K, Rb, Cs) exhibit instability from weak bonding and lattice softening, and certain transition metals (V, Mo, W) are prone to unstable configurations due to complex coordination. These results highlight the chemical dependence of dynamical instability in AI-generated structures.}}
		\label{Element} 
\end{figure}

\textcolor{black}{To further investigate the origins of dynamical instability in AI-generated structures, we performed an element-level statistical analysis. Specifically, each structure was decomposed into its constituent elements, and, at a given imaginary frequency threshold ($-0.001$-$-1.0$~THz,Figure~\ref{Element}) , we calculated the fraction of structures containing a given element that exhibited dynamical instability (i.e., possessed imaginary phonon modes), yielding an element-resolved instability probability ranking. The results indicate that the elements with the highest instability probabilities are mainly nonmetals such as F, O, Cl, H, Br, and I (all exceeding 85\%), followed by transition metals such as W, V, Mo, Nb, and alkali metals including K, Rb, and Cs. This analysis demonstrates that dynamical instability exhibits a strong chemical dependence rather than occurring randomly. From a physical perspective, highly electronegative nonmetals such as F, O, and Cl form localized, highly directional, and rigid bonds that are sensitive to small variations in bond lengths and angles; even minor geometric distortions in the generated structures can induce negative eigenvalues in the force constant matrix, producing imaginary modes. For alkali metals like K, Rb, and Cs, their large ionic radii, weak bonding with anions, and low electronegativity render the lattice “soft,” readily introducing low-frequency unstable modes. In the case of transition metals such as V, Mo, and W, instability arises from more complex mechanisms, often linked to variable coordination environments and the involvement of $d$-electrons in bonding. These physical factors make the structures highly sensitive to minor atomic perturbations, resulting in AI-generated configurations that may appear energetically reasonable yet are dynamically unstable. Overall, these results demonstrate that dynamical instability has a clear chemical origin, reflecting systematic challenges of current generative models in capturing strong bonding constraints, lattice flexibility, and complex coordination environments, rather than arising from random prediction errors.}

\subsection{\textcolor{black}{Chemical-Complexity}}
\begin{figure}[H]
		\centering  
		\includegraphics[width=1.0\linewidth]{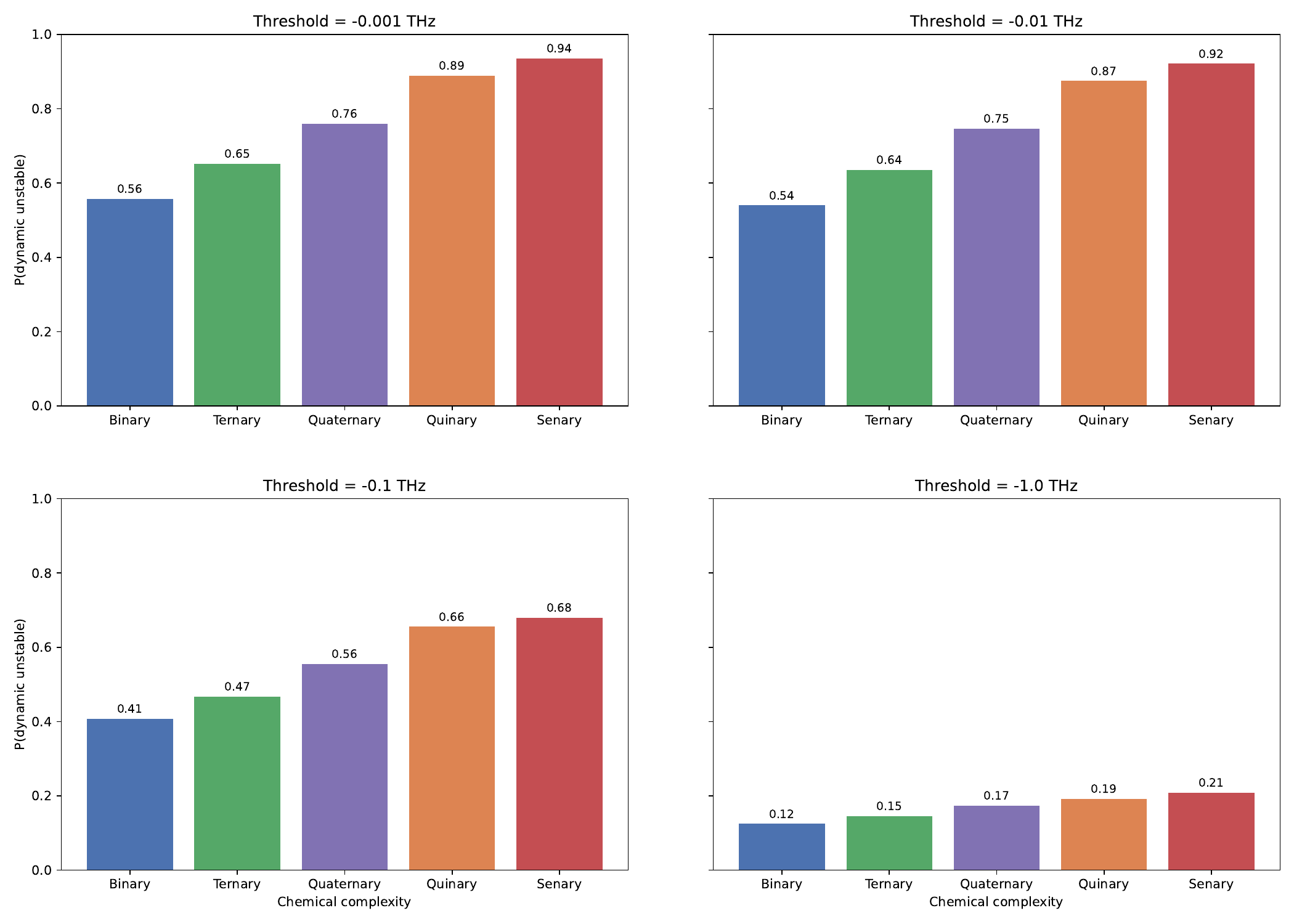}
        \color{black}
\caption{\textcolor{black}{Dynamical instability as a function of chemical complexity. Structures are grouped by the number of constituent elements (binary to senary), and instability probabilities are shown under four imaginary frequency thresholds ($-0.001$, $-0.01$, $-0.1$, $-1.0$~THz). The monotonic increase in instability with chemical complexity highlights that multicomponent systems are intrinsically more prone to dynamical instabilities.}}
		\label{chemical-complexity} 
\end{figure}

\textcolor{black}{To further investigate the physical origin of dynamical instability, we performed a statistical analysis from the perspective of chemical complexity. As show Figure~\ref{chemical-complexity}, each structure was categorized based on the number of constituent elements (binary, ternary, quaternary, quinary, and senary systems), and for each category, we computed the probability of dynamical instability (i.e., the presence of imaginary phonon frequencies) under different frequency thresholds. As shown by the results, a clear and monotonic trend emerges across all thresholds: the probability of dynamical instability systematically increases with chemical complexity. For instance, under the strict threshold of $-0.001$~THz, the instability probability rises from 55.78\% for binary systems to 93.56\% for senary systems. A similar trend is consistently observed at other thresholds, such as $-0.1$~THz (from 40.73\% to 67.93\%) and $-1.0$~THz (from 12.47\% to 20.87\%), demonstrating that this behavior is robust and not sensitive to the specific choice of threshold. This systematic increase can be attributed to the growing structural and chemical complexity in multicomponent systems. As the number of constituent elements increases, the configurational space expands rapidly, making it more difficult to simultaneously satisfy local bonding preferences, charge balance, and geometric constraints. This often leads to local distortions, lattice frustration, and soft phonon modes, which manifest as dynamical instabilities. In contrast, binary systems have more constrained and well-defined bonding environments, making them comparatively easier to stabilize. These results reveal that dynamical instability is strongly correlated with chemical complexity, highlighting an intrinsic limitation of current generative models: while they can explore vast compositional spaces, they face increasing difficulty in producing dynamically stable structures as the number of constituent elements grows. This finding provides direct physical insight into the failure modes of AI-generated materials beyond simple performance metrics.}

\subsection{\textcolor{black}{Coordination Environments}}
\textcolor{black}{We systematically analyze the relationship between coordination environments and dynamical instability by binning structures according to their average coordination number ($\mathrm{avg\_cn}$).The coordination number (CN) of atom $i$ is defined as the number of neighboring atoms within a given distance range:
\begin{equation}
\mathrm{CN}_i = \sum_{j \neq i} f(r_{ij}),
\end{equation}
where $r_{ij}$ is the distance between atoms $i$ and $j$, and $f(r_{ij})$ is a neighbor indicator function:
\begin{equation}
f(r_{ij}) =
\begin{cases}
1, & r_{ij} \le r_\mathrm{cut} \\
0, & r_{ij} > r_\mathrm{cut}
\end{cases}.
\end{equation}
In this work, the function $f(r_{ij})$ is determined adaptively using the \texttt{MinimumDistanceNN} method from \texttt{pymatgen}, which identifies chemically reasonable nearest neighbors for each atom based on interatomic distances.}

\begin{figure}[H]
		\centering  
		\includegraphics[width=1.0\linewidth]{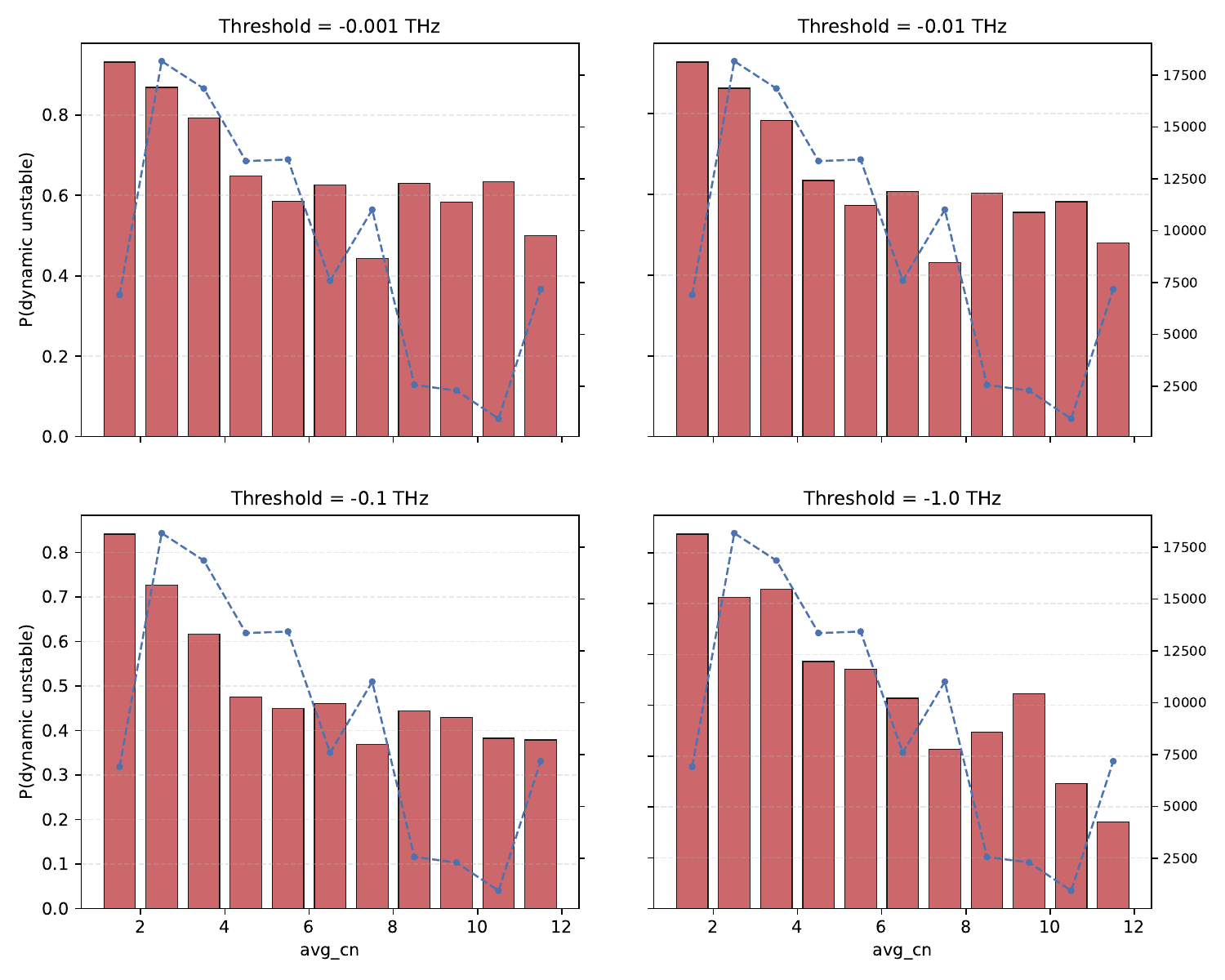}
        \color{black}
\caption{
\textcolor{black}{Dynamical instability probability as a function of average coordination number ($\mathrm{avg\_cn}$) under different imaginary-frequency thresholds ($-0.001$, $-0.01$, $-0.1$, and $-1.0$~THz). The coordination number is computed using a distance-based nearest-neighbor criterion (MinimumDistanceNN), reflecting local atomic connectivity. A clear and consistent trend is observed across all thresholds: structures with lower coordination numbers exhibit significantly higher instability probabilities, while higher coordination leads to enhanced lattice stability.}
}
		\label{CN-com} 
\end{figure}

\textcolor{black}{The probability of dynamical instability is defined as the fraction of structures exhibiting imaginary phonon modes below a given frequency threshold (from $-0.001$ to $-1.0$~THz). Our results reveal a consistent and pronounced coordination dependence across all thresholds. As shown in Figure~\ref{CN-com}, the instability probability decreases from approximately $84\%$ in the low-coordination regime ($\mathrm{avg\_cn} \approx 1.5$) to about $45\%$ at intermediate coordination ($\mathrm{avg\_cn} \approx 5.5$), and further to $\sim 38\%$ in the high-coordination regime ($\mathrm{avg\_cn} \gtrsim 9$), exhibiting an overall monotonic trend. This indicates that under-coordinated structures, lacking sufficient geometric constraints, are highly prone to soft phonon modes, whereas increasing coordination enhances lattice connectivity and mechanical rigidity, thereby suppressing dynamical instabilities. This trend becomes even more pronounced under stricter criteria ($-0.001$~THz), where low-coordination structures exhibit instability probabilities as high as $\sim 93\%$, while intermediate-to-high coordination structures reduce to $\sim 45$--$65\%$. Under a more relaxed threshold ($-1.0$~THz), although the overall instability probability decreases significantly (e.g., from $\sim 37\%$ to $\sim 8\%$ as $\mathrm{avg\_cn}$ increases from $1.5$ to $11.5$), the negative correlation between coordination number and instability remains robust. This demonstrates that the observed trend is not an artifact of threshold selection but reflects an intrinsic structural property. Overall, the average coordination number serves as an effective descriptor for distinguishing stability regimes and reveals a general physical mechanism: local atomic connectivity governs lattice rigidity and thereby controls the emergence of soft phonon modes.}

\subsection{\textcolor{black}{Formation Energy}}
 \textcolor{black}{Dynamical instability can arise from multiple factors. From a physical perspective, thermodynamic stability and dynamical stability describe complementary aspects of material stability: the former reflects the global energetic favorability against decomposition, while the latter characterizes the local curvature of the potential energy surface around a given structure. One intuitive mechanism linking the two is thermodynamic instability. Structures that are energetically unfavorable (e.g., with high formation energy or lying above the convex hull) are more likely to undergo structural distortions or decomposition, which may manifest as imaginary phonon modes. The formation energy is defined as
\begin{equation}
E_{\mathrm{form}} = \frac{1}{N} \left( E_{\mathrm{tot}} - \sum_i n_i \mu_i \right),
\end{equation}
where $E_{\mathrm{tot}}$ is the total energy of the compound, $n_i$ is the number of atoms of element $i$, and $\mu_i$ is the chemical potential of the corresponding reference state.}

\begin{figure}[H]
		\centering  
		\includegraphics[width=0.9\linewidth]{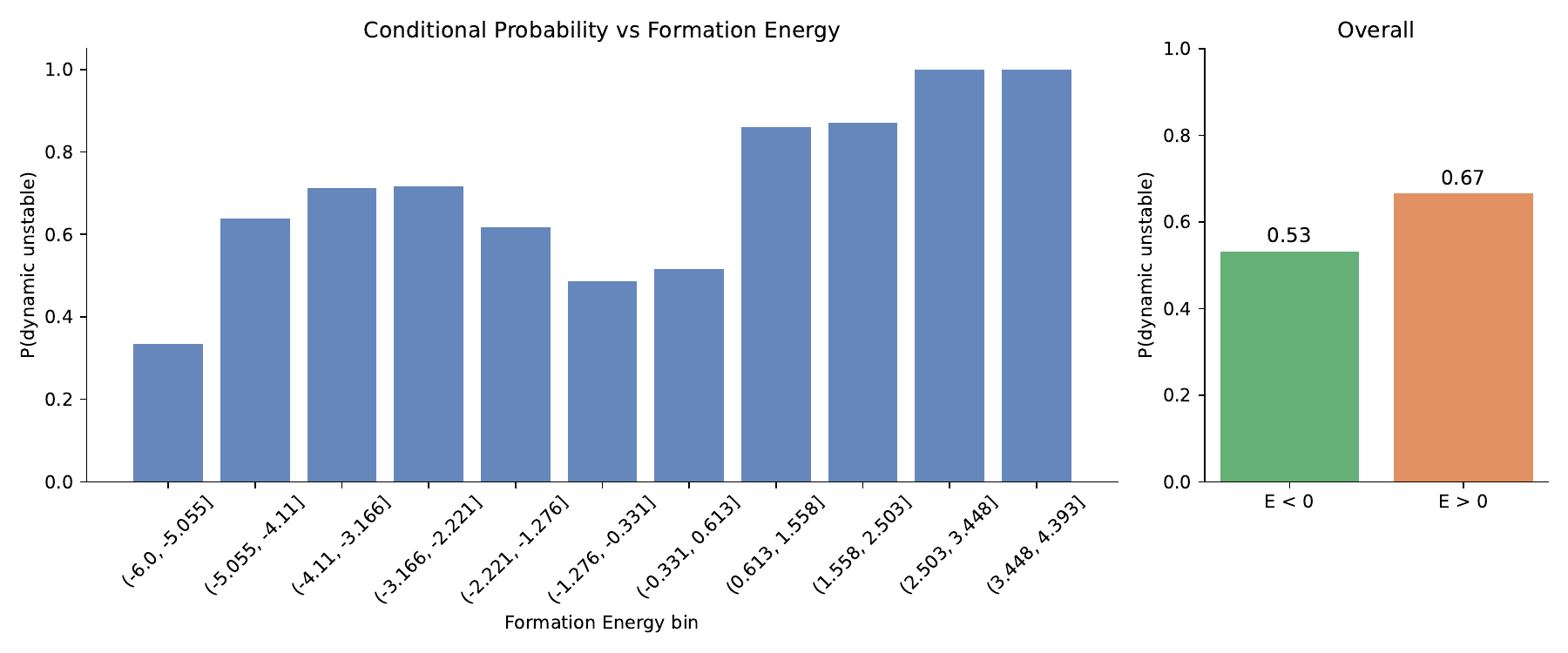}
        \color{black}
		\caption{\textcolor{black}{Formation energy versus dynamical stability. Thermodynamically unstable materials ($E_{\mathrm{form}} > 0$) show a moderately higher probability of dynamical instability; however, the weak overall correlation (Pearson $r = 0.0133$) indicates that formation energy is not a reliable predictor of dynamical stability.}}
		\label{formation-phonon} 
\end{figure}

 \textcolor{black}{As shown in Figure~\ref{formation-phonon} and Table~\ref{avg-cn-un}, we perform a statistical analysis over all materials. Under an imaginary frequency threshold of $-0.1$~THz, 66.62\% of materials with positive formation energy are dynamically unstable, compared to 53.21\% for those with negative formation energy. This indicates that thermodynamic instability increases the likelihood of dynamical instability, suggesting a certain degree of correlation. However, the increase is modest (approximately 14\%), implying that the effect is limited. From a global statistical perspective, the correlation remains very weak. Specifically, the Pearson correlation coefficient between formation energy and the minimum phonon frequency is only 0.0133, which is close to zero, indicating no significant linear relationship. The formation energies used in this analysis are obtained from a MEGNet model trained on the GNoME dataset, with a reported prediction error of approximately 0.0021~eV. Given this high level of accuracy, the observed weak correlation cannot be attributed to prediction noise, but instead reflects an intrinsic decoupling between thermodynamic and dynamical stability.}

\begin{table}[H]
\centering
\color{black}
\caption{\textcolor{black}{Binned dynamical instability probability ($P_{\mathrm{unstable}}$) as a function of average coordination number ($\mathrm{avg\_cn}$) under different imaginary-frequency thresholds.}}
\label{tab:cn_instability}
\begin{tabular}{ccc|ccc}
\toprule
\multicolumn{3}{c|}{$-0.001$ THz} & \multicolumn{3}{c}{$-0.010$ THz} \\
\midrule
Center & $P_{\mathrm{unstable}}$ & Count & Center & $P_{\mathrm{unstable}}$ & Count \\
\midrule
1.5  & 0.9317 & 6912  & 1.5  & 0.9280 & 6912 \\
2.5  & 0.8684 & 18180 & 2.5  & 0.8629 & 18180 \\
3.5  & 0.7925 & 16859 & 3.5  & 0.7837 & 16859 \\
4.5  & 0.6479 & 13362 & 4.5  & 0.6343 & 13362 \\
5.5  & 0.5860 & 13434 & 5.5  & 0.5733 & 13434 \\
6.5  & 0.6257 & 7595  & 6.5  & 0.6074 & 7595  \\
7.5  & 0.4425 & 11015 & 7.5  & 0.4308 & 11015 \\
8.5  & 0.6305 & 2563  & 8.5  & 0.6028 & 2563  \\
9.5  & 0.5840 & 2293  & 9.5  & 0.5556 & 2293  \\
10.5 & 0.6334 & 933   & 10.5 & 0.5820 & 933   \\
11.5 & 0.4998 & 7179  & 11.5 & 0.4793 & 7179  \\
\midrule
\multicolumn{3}{c|}{$-0.100$ THz} & \multicolumn{3}{c}{$-1.000$ THz} \\
\midrule
Center & $P_{\mathrm{unstable}}$ & Count & Center & $P_{\mathrm{unstable}}$ & Count \\
\midrule
1.5  & 0.8413 & 6912  & 1.5  & 0.3685 & 6912  \\
2.5  & 0.7258 & 18180 & 2.5  & 0.3059 & 18180 \\
3.5  & 0.6160 & 16859 & 3.5  & 0.3141 & 16859 \\
4.5  & 0.4755 & 13362 & 4.5  & 0.2432 & 13362 \\
5.5  & 0.4491 & 13434 & 5.5  & 0.2357 & 13434 \\
6.5  & 0.4606 & 7595  & 6.5  & 0.2068 & 7595  \\
7.5  & 0.3682 & 11015 & 7.5  & 0.1569 & 11015 \\
8.5  & 0.4436 & 2563  & 8.5  & 0.1736 & 2563  \\
9.5  & 0.4300 & 2293  & 9.5  & 0.2115 & 2293  \\
10.5 & 0.3826 & 933   & 10.5 & 0.1233 & 933   \\
11.5 & 0.3787 & 7179  & 11.5 & 0.0852 & 7179  \\
\bottomrule
\end{tabular}
\label{avg-cn-un}
\end{table}

\subsection{\textcolor{black}{Metal vs.\ Insulator}}
\begin{figure}[H]
		\centering  
		\includegraphics[width=0.65\linewidth]{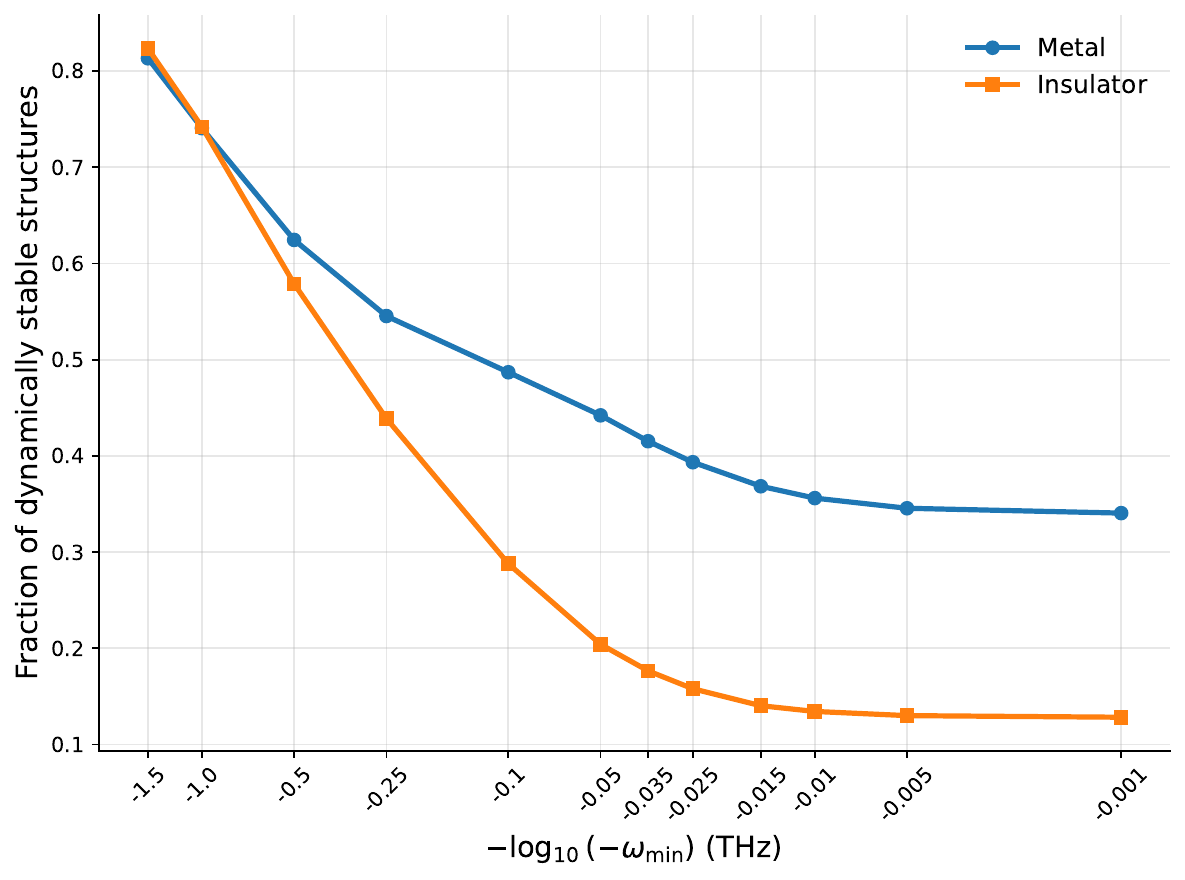}
        \color{black}
		\caption{\textcolor{black}{Fraction of dynamically stable structures versus the minimum phonon frequency threshold for metallic and insulating systems. Insulators show a higher proportion of unstable structures in the weak instability regime (e.g., $-0.01$ to $-0.1$~THz), whereas the difference between metals and insulators becomes negligible at stronger instability thresholds. This trend suggests that the excess instability in insulators is primarily associated with soft phonon modes.}}
		\label{metal-ins} 
\end{figure}
\textcolor{black}{To further understand the origin of the differences in dynamical stability among crystal generation models, we analyze the problem from the perspective of electronic character (metal vs.\ insulator). Geometric graph neural network (GNN)-based models have demonstrated strong performance in predicting material properties, particularly formation energy and band gap. Based on this, we train an ALIGNN-based~\cite{Choudhary2021} classifier to distinguish metals from insulators, achieving an accuracy of 94\% on the Materials Project dataset. This classifier is then applied to all generated structures to assign their electronic character. We subsequently evaluate the fraction of dynamically unstable structures in metallic and insulating systems as a function of the instability threshold defined by the minimum phonon frequency. As shown in Figure~\ref{metal-ins}, under weak instability thresholds (e.g., $-0.01$ to $-0.1$~THz), the fraction of unstable structures is significantly higher in insulators than in metals. However, as the threshold decreases (corresponding to stronger instabilities), the difference between the two classes rapidly diminishes. When the threshold is below approximately $-1$~THz, the instability fractions of metals and insulators become nearly identical. This trend suggests that the excess instability observed in insulating systems is primarily concentrated in the regime of weak soft phonon modes. Such instabilities are typically associated with low-energy structural distortions, such as ferroelectric instabilities, octahedral tilting, or symmetry breaking, which are more common in insulating materials. In contrast, strong instabilities (characterized by large imaginary frequencies) are more likely to originate from unrealistic local geometries or insufficient structural relaxation, and thus affect metallic and insulating systems in a similar manner.}

\subsection{\textcolor{black}{Space Group Symmetry}}
\begin{figure}[H]
		\centering  
		\includegraphics[width=0.65\linewidth]{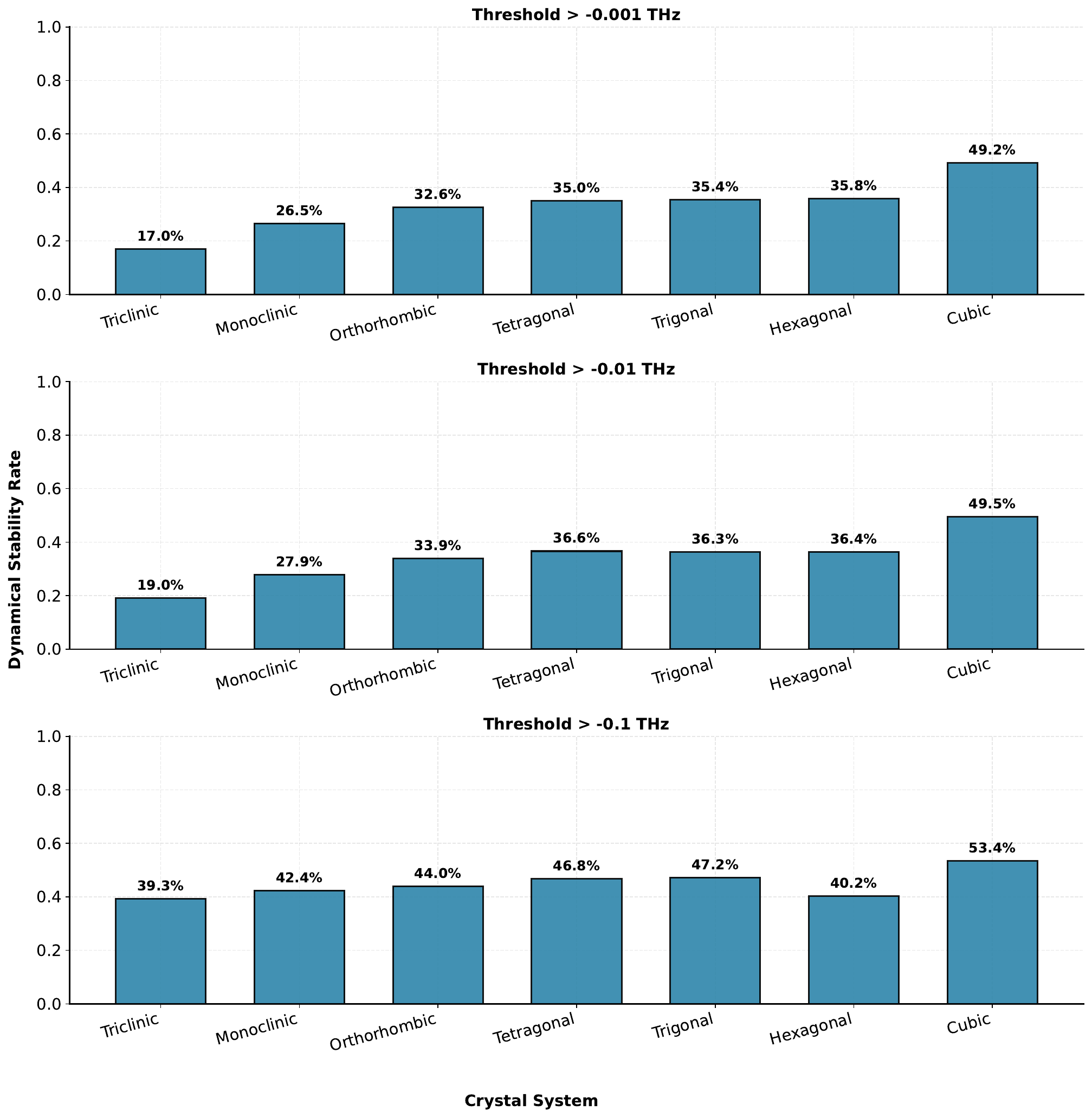}
        \color{black}
		\caption{\textcolor{black}{Dynamical stability ratios of the seven crystal systems under different imaginary phonon frequency thresholds ($-0.001$, $-0.01$, and $-0.1$~THz). The bar chart shows the fraction of dynamically stable structures within each crystal system. Across all thresholds, higher-symmetry crystal systems (e.g., cubic and trigonal) consistently exhibit higher stability ratios than lower-symmetry systems (e.g., triclinic and monoclinic), and the relative ordering remains largely unchanged as the threshold is relaxed.}}
		\label{fig-multi-threshold-crystal-sys} 
\end{figure}

\textcolor{black}{As shown in Figure~\ref{fig-multi-threshold-crystal-sys}, under different imaginary-frequency thresholds ($-0.001$, $-0.01$, and $-0.1$~THz), the dynamical stability of various crystal systems exhibits consistent and physically meaningful systematic trends.}

\textcolor{black}{Under the strictest threshold ($-0.001$~THz), the overall stability ratios are relatively low, with pronounced differences across crystal systems: triclinic structures show a stability of only 17.0\%, whereas cubic systems reach 49.2\%. A clear monotonic trend is observed with increasing crystal symmetry, from low-symmetry systems (triclinic and monoclinic), through intermediate-symmetry systems (orthorhombic and tetragonal), to high-symmetry systems (trigonal, hexagonal, and cubic), with stability progressively increasing. When the threshold is relaxed to $-0.01$~THz, the stability ratios of all systems increase slightly, while the overall ranking remains nearly unchanged. This indicates that the differences in stability are not driven by a few extreme imaginary modes but instead originate from intrinsic structural characteristics. Upon further relaxation to $-0.1$~THz, the stability ratios increase significantly—for example, triclinic systems rise from 17.0\% to 39.3\%, orthorhombic systems from 32.6\% to 44.0\%, and cubic systems from 49.2\% to 53.4\%. Despite this overall increase, the relative ordering across crystal systems remains consistent, with higher-symmetry crystals consistently exhibiting greater dynamical stability. This cross-threshold consistency demonstrates the robustness of the observed trend and confirms that it is not an artifact of statistical fluctuations or threshold selection.}



\textcolor{black}{This behavior originates from symmetry constraints imposed by Neumann’s principle~\cite{nye1985physical, tinkham2003group,born1996dynamical}, which requires that any physical property tensor, including the force constant (Hessian) matrix, remains invariant under all symmetry operations of the crystal. Mathematically, for any symmetry operation $R$ of the space group, the Hessian matrix $\Phi$ satisfies the invariance condition
\begin{equation}
\Phi = R \, \Phi \, R^{T},
\end{equation}
which, in component form, can be written as
\begin{equation}
\Phi_{i\alpha, j\beta} = \sum_{\alpha'\beta'} R_{\alpha\alpha'} \, R_{\beta\beta'} \, \Phi_{i'\alpha', j'\beta'}.
\end{equation}
Here, $i,j$ label atoms and $\alpha,\beta$ denote Cartesian components. This constraint enforces strict symmetry relations among matrix elements, significantly reducing the number of independent degrees of freedom and restricting the admissible form of the Hessian matrix~\cite{nye1985physical, tinkham2003group}.}

\textcolor{black}{From a lattice dynamical perspective, phonon frequencies are obtained from the eigenvalue problem of the dynamical matrix $D(\mathbf{q})$, defined as
\begin{equation}
D_{i\alpha,j\beta}(\mathbf{q}) = \frac{1}{\sqrt{m_i m_j}} \sum_{\mathbf{R}} \Phi_{i\alpha,j\beta}(\mathbf{R}) e^{i \mathbf{q} \cdot \mathbf{R}},
\end{equation}
where $m_i$ is the atomic mass and $\mathbf{q}$ is the phonon wavevector. The phonon frequencies $\omega$ are determined by
\begin{equation}
\det \left| D(\mathbf{q}) - \omega^2 I \right| = 0.
\end{equation}
Symmetry constraints on $\Phi$ directly propagate to $D(\mathbf{q})$, enforcing block-diagonal structures according to irreducible representations of the space group and restricting the distribution of eigenvalues.}

\textcolor{black}{Within our benchmark dataset, a robust empirical correlation is observed between higher structural symmetry and the presence of fewer imaginary phonon modes; this relationship is consistent with the symmetry-imposed constraints on the force constant and dynamical matrices. In contrast, lower-symmetry systems possess more unconstrained degrees of freedom, making them more susceptible to local distortions and soft vibrational modes. Therefore, high-symmetry crystals (e.g., cubic and hexagonal systems) are statistically less likely to exhibit imaginary modes and thus display higher dynamical stability.}

\textcolor{black}{Importantly, the consistency of this symmetry–stability relationship across different frequency thresholds and crystal systems in our benchmark demonstrates that AI-generated structures also obey these fundamental physical constraints. This indicates that the observed trends are not artifacts of specific models or datasets, but rather reflect intrinsic symmetry-governed properties of crystal lattices.}

\section{\textcolor{black}{Effect of Training Data Scale on Dynamical Stability}}
\label{app:scale}
\begin{figure}[H]
		\centering  
		\includegraphics[width=0.65\linewidth]{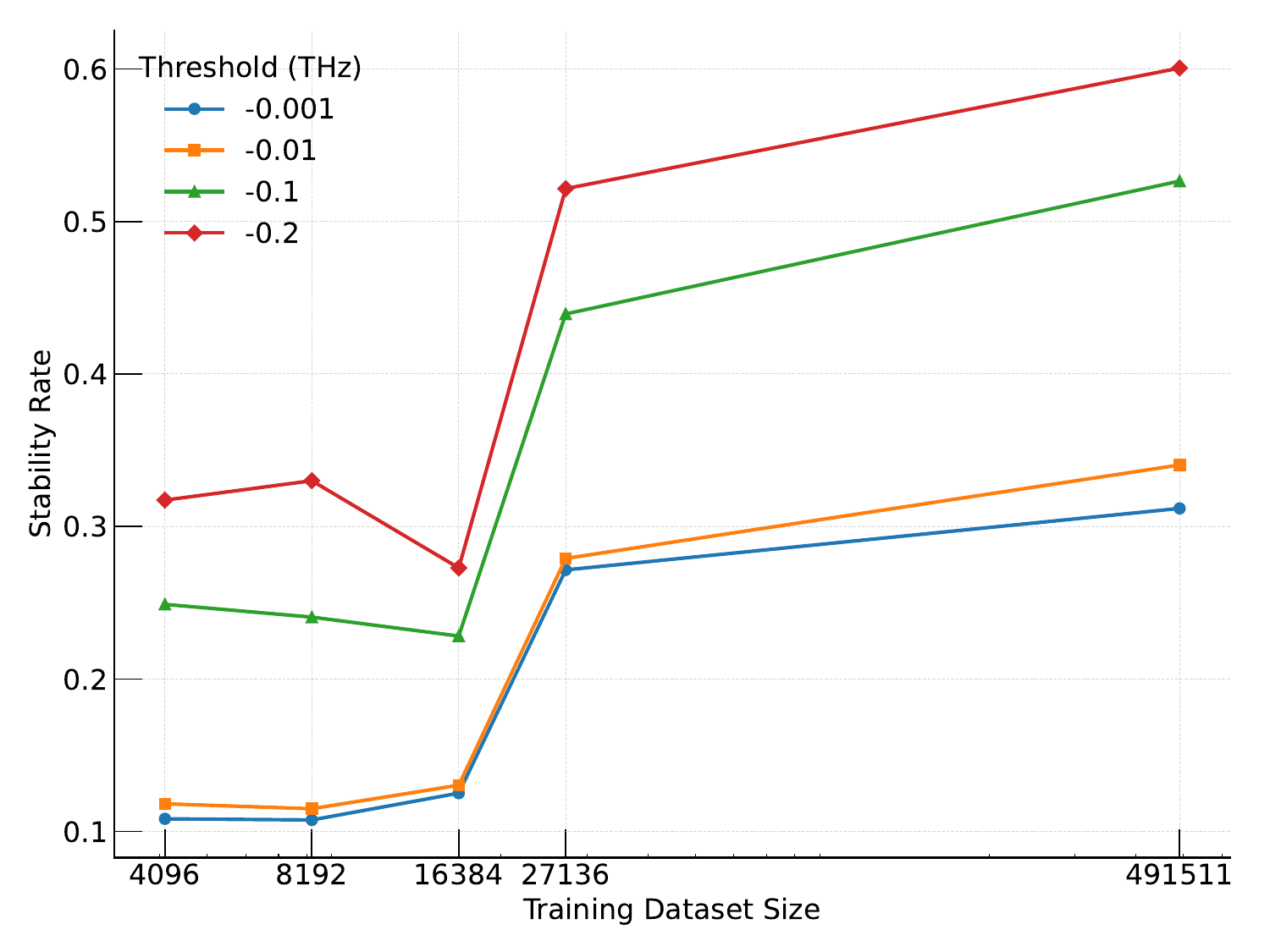}
        \color{black}
		\caption{\textcolor{black}{Scaling behavior of phonon dynamical stability with respect to training dataset size for DiffCSP models. Models are trained on datasets ranging from 4,096 to 491,511 structures, and stability is evaluated under different imaginary frequency thresholds (-0.001, -0.01, -0.1, and -0.2 THz).}}
		\label{fig-stability_scaling} 
\end{figure}

\textcolor{black}{To systematically investigate the effect of training data scale on the dynamical stability of generated materials, we constructed a series of models based on the DiffCSP framework, with training set sizes of 4,096, 8,192, 16,384, 27,136, and 491,511, respectively. For each model, a set of candidate crystal structures was generated and deduplicated against the corresponding training set to eliminate potential memorization effects (Table~\ref{tab:scaling_stability} and Figure~\ref{fig-stability_scaling}). Phonon calculations were then performed using MatterSim, with more than 3,500 structures retained for each model to ensure statistical reliability and comparability. The results show that under strict dynamical stability criteria (e.g., imaginary frequency thresholds of $-0.001$ and $-0.01$ THz), the stability of generated structures improves significantly with increasing training set size. Specifically, the stability rate increases from approximately 0.108 to 0.312 at $-0.001$ THz, and from 0.118 to 0.340 at $-0.01$ THz, indicating that larger datasets substantially enhance the model’s ability to learn stable configurations. However, under more relaxed criteria (e.g., $-0.1$ and $-0.2$ THz), the stability does not increase monotonically; instead, it exhibits a slight decrease or fluctuation before rising again. For instance, at the $-0.1$ THz threshold, the stability decreases from 0.249 to 0.228 (at 16,384), followed by a marked increase to 0.527 at larger scales. This behavior suggests that, at intermediate data scales, the model tends to generate a significant number of structures near the boundary of dynamical stability, i.e., “marginally unstable” materials with small imaginary frequencies. Such behavior is consistent with the well-known soft-mode instabilities encountered in first-principles calculations and reflects a longstanding challenge in DFT-based studies. As the training data scale further increases, the proportion of these borderline unstable structures decreases, and the model becomes more capable of distinguishing stable from unstable regions, leading to a substantial improvement in overall dynamical stability. This trend is also observed under more relaxed thresholds (e.g., $-0.4$ to $-1.0$ THz), further highlighting the critical role of large-scale data in improving the physical reliability of generative models.}

\textcolor{black}{While it is generally expected that diffusion-based generative models and pretraining on large-scale datasets (e.g., Alex20) can improve stability, our study goes beyond qualitative observations by providing a systematic and quantitative analysis of how dynamical stability evolves with training data scale. In particular, we explicitly characterize stability trends across multiple dataset sizes (Figure~\ref{fig-stability_scaling}), offering empirical insights into the scaling behavior of generative models. At the same time, our results reveal that this improvement is not unbounded. Increasing the dataset size does not lead to a proportional or monotonic gain in phonon stability, and the performance gain gradually saturates at larger scales. Moreover, we observe that generative models, similar to DFT-based human design, encounter the common challenge of producing structures near the boundary of dynamical stability. These findings suggest that simply scaling up data alone is insufficient to fundamentally resolve stability issues. Instead, they highlight the need for incorporating more explicit physical constraints or inductive biases into generative models to achieve further improvements.}

\begin{table}[H]
\centering
\color{black}
\caption{\textcolor{black}{Phonon stability of structures generated by DiffCSP models trained on datasets of varying sizes. Stability is evaluated under different imaginary frequency thresholds (THz).}}
\label{tab:scaling_stability}
\begin{tabular}{c|ccccccc|c}
\hline
\textbf{Dataset Size} & \textbf{-0.001} & \textbf{-0.01} & \textbf{-0.1} & \textbf{-0.2} & \textbf{-0.4} & \textbf{-0.8} & \textbf{-1.0} & \textbf{Mean $\omega_{\min}$}  \\
\hline
4096   & 0.108 & 0.118 & 0.249 & 0.317 & 0.404 & 0.529 & 0.581 & -2.47  \\
8192   & 0.108 & 0.115 & 0.241 & 0.330 & 0.432 & 0.603 & 0.665 & -1.16  \\
16384  & 0.125 & 0.130 & 0.228 & 0.273 & 0.323 & 0.404 & 0.436 & -2.14  \\
27136  & 0.272 & 0.279 & 0.439 & 0.522 & 0.617 & 0.753 & 0.803 & -0.64  \\
491511 & 0.312 & 0.340 & 0.527 & 0.601 & 0.710 & 0.838 & 0.878 & -0.46 \\
\hline
\end{tabular}
\end{table}

\section{\textcolor{black}{Enhancing LLM-Based Crystal Generation through Prompting and Scaling}}
\label{app:llm}
\subsection{\textcolor{black}{CrystaLLM-large}}
\textcolor{black}{First, in the CrystaLLM experiments, we use the smaller models and the larger and more powerful \texttt{crystallm\_mp\_20\_large} model. Trained on the \texttt{MP20} dataset, this model has significantly higher structure-generation capability compared to earlier versions (\texttt{crystallm\_mp\_20\_small}), allowing for a more objective assessment of CrystaLLM’s performance in crystal generation tasks.}

\textcolor{black}{Second, regarding prompt design, CrystaLLM supports multiple conditional prompting strategies, with the most common being \emph{chemical formula prompting} and \emph{chemical formula + space group prompting}. In this study, we applied both strategies to perform large-scale structure generation experiments. Specifically, for each prompting strategy, 10,902 crystal structures were generated, yielding a total of 21,804 candidate structures. It is important to note that these prompts were not taken directly from the \texttt{Materials Project} database but were derived from new candidate materials generated by \texttt{MatterGen}. Since our goal is to evaluate the model’s ability to generate stable structures in unexplored chemical space rather than reproduce known database entries, we extracted chemical formulas and space group information from the \texttt{MatterGen} candidates to serve as generation conditions, thereby constructing a more exploration-oriented task.}
\textcolor{black}{Finally, following structure generation, we applied the official CrystaLLM postprocessing workflow. The provided postprocess.py script was used to standardize the generated structures, including correcting symmetry operators and removing nonstandard atomic attributes to ensure compliance with CIF format specifications. Subsequently, all generated structures were subjected to structural relaxation and phonon calculations using MatterSim, allowing a systematic assessment of the dynamical stability of the generated materials. In the final evaluation, we followed the reviewer’s suggestion and applied three commonly used imaginary frequency thresholds (-0.001 THz,-0.01 THz, -0.1 THz, and -1 THz) to statistically analyze the stability of the generated materials, with the corresponding results reported in the Table~\ref{tab:stability_grouped}.}

\subsection{\textcolor{black}{LLaMA-2 70B}}
\textcolor{black}{We additionally evaluated another class of LLM-based crystal generation methods, namely the \textbf{Crystal-Text-LLM}~\cite{iclr2024LLM} framework (titled ``\emph{Fine-Tuned Language Models Generate Stable Inorganic Materials as Text}, ICLR 2024''). This framework generates crystal structures by fine-tuning a pre-trained large language model on text-encoded crystal representations. Among the models within this framework, \texttt{LLaMA-2 70B} exhibited the best performance and was fine-tuned on the \texttt{MP-20} dataset.  For this model, we directly used the initial crystal generation data publicly released on GitHub~\footnote{\url{https://github.com/facebookresearch/crystal-text-llm/tree/main/saved_samples}}. In contrast to the single CrystaLLM model criticized for suboptimal prompting, using LLaMA2-70B ensures that our evaluation reflects the capabilities of well-configured LLM-based generative methods rather than a single poorly prompted instance. The dataset includes four different sampling parameter settings (\texttt{llama-2-70B\_0.7\_0.7.csv},\\  \texttt{llama-2-70B\_0.7\_1.0.csv}, \texttt{llama-2-70B\_1.0\_0.7.csv}, and \texttt{llama-2-70B\_1.0\_1.0.csv}), totaling 4 × 12,500 entries. After applying a similar deduplication procedure, we randomly selected 1,250 materials from each parameter set, resulting in 5,000 structures for phonon calculations. Ultimately, 4,633 crystals successfully completed structural relaxation, and the corresponding phonon calculation results are presented in Table~\ref{tab:LLM70B}.}

\textcolor{black}{From the results, it is evident that the larger \texttt{CrystaLLM} models exhibit substantially improved stability compared to the small, unprompted version. For example, under the strictest phonon threshold of $-0.001$~THz, the two large \texttt{CrystaLLM} models achieve stability rates of \textbf{18.4\%} and \textbf{19.6\%}, exceeding both \texttt{CrystalFlow} (16.8\%) and \texttt{CrystalFormer} (11.6\%) trained on MP-20. This demonstrates that appropriate scaling and prompt design significantly enhance LLM-based crystal generation performance. Moreover, the \texttt{LLaMA2-70B} model further improves stability, achieving \textbf{21.7\%} under the $-0.001$~THz threshold, surpassing both large \texttt{CrystaLLM} models as well as the traditional generators mentioned above. Table~\ref{tab:LLM70B} further shows that this performance is robust across different sampling parameters (temperature $\tau$ and top-$p$), with stability consistently exceeding 19\% and reaching up to \textbf{23.9\%} under certain parameter choices. These results indicate that optimized prompting and parameter tuning can substantially improve the dynamical stability of generated structures. Taken together, these findings support a more nuanced conclusion: while early LLM-based models may have lagged behind GNN-based approaches, larger LLM models with carefully designed prompts and optimized sampling can achieve stability levels comparable to, and in some cases exceeding, traditional crystal generation methods. We have updated the revised manuscript to reflect this broader and more accurate assessment, avoiding overly general claims regarding architectural limitations.}

\textcolor{black}{However, we emphasize that these improvements should be interpreted in the context of fundamental differences between the two paradigms. LLMbased Generators typically contain significantly more parameters than GNN-based Generators, often by several orders of magnitude, resulting in substantially higher computational requirements. In addition, LLM-based crystal generation critically depends on prompt design and sampling strategies, which introduce additional degrees of freedom and require careful tuning. In contrast, GNN-based generative models are inherently structured for atomistic systems and do not rely on external prompts, making them more straightforward to apply in practice.}

\begin{table}[H]
\centering
\color{black}
\caption{Dynamical stability of different crystal generation models. 
Models are grouped by type: traditional crystal generators and LLM/Transformer-based generators. 
Stability is defined using the minimum phonon frequency threshold (THz). Notes: "MP20" and "Alex20" indicate the training datasets. For LLM/Transformer models, CrystaLLM (small, no-prompt, MP20) is a small model without any prompt, whereas CrystaLLM (large, formula, MP20) and CrystaLLM (large, formula+sg, MP20) are large models using chemical formula prompts, with the latter also including space group information. LLaMA2-70B (70B, optimal, MP20) is a large model employing the author-optimized prompt. "Relaxed" refers to the number of structures successfully geometry-relaxed, and the stability columns show the fraction of dynamically stable structures under different phonon thresholds.}
\begin{tabular}{lccccc}
\hline
\multirow{2}{*}{Model} & \multirow{2}{*}{Relaxed} & \multicolumn{4}{c}{Frequency Threshold (THz)} \\
\cline{3-6}
 &  & $>-0.001$ & $>-0.01$ & $>-0.1$ & $>-1.0$ \\
\hline
\multicolumn{6}{c}{\textbf{GNN-based Generators}} \\
\hline
CrystalFlow (MP20)           & 8533   & 0.1676 & 0.1763 & 0.4007 & \textbf{0.8260} \\
CrystalFormer (MP20)             & 4408   & 0.1157 & 0.1198 & 0.1554 & 0.3058 \\
DiffCSP (MP20)                   & 9163   & \textbf{0.2715} & \textbf{0.2791} & 0.4394 & 0.8030 \\
InvDesFlow-AL (MP20) & 8000 & 0.2702 & 0.2778 & 0.4378  & 0.8037\\
MatterGen (MP20)                 & 9279   & 0.2455 & 0.2557 & \textbf{0.4505} & 0.7915 \\
\hline
InvDesFlow-AL (Alex20)           & 22755  & 0.3119 & 0.3403 & 0.5265 & 0.8775 \\
CrystalFormer (Alex20)           & 8642   & 0.3436 & 0.3555 & 0.4539 & 0.7234 \\
MatterGen (Alex20)               & 10902  & \textbf{0.4099} & \textbf{0.4295} & \textbf{0.5789} & \textbf{0.8818} \\
\hline
\multicolumn{6}{c}{\textbf{LLM / Transformer-based Generators}} \\
\hline
CrystaLLM (small, no-prompt,MP20)         & 1951   & 0.0297 & 0.0369 & 0.1435 & 0.5864 \\
CrystaLLM (large, formula, MP20) & 10387  & 0.1840 & 0.1880 & 0.2340 & 0.4750 \\
CrystaLLM (large, formula+sg, MP20) & 9975  & 0.1960 & 0.2030 & 0.3110 & \textbf{0.5930} \\
LLaMA2-70B (70B,optimal,MP20)               & 4633   & \textbf{0.2170} & \textbf{0.2250} & \textbf{0.3250} & 0.5780 \\
\hline
\end{tabular}
\label{tab:stability_grouped}
\end{table}

\begin{table}[H]
\centering
\color{black}
\caption{Dynamical stability rates of LLaMA2-70B (70B, optimal, MP-20) under different sampling parameters (temperature $\tau$ and top-$p$). Stability is evaluated using the minimum phonon frequency threshold (THz).}
\label{tab:sampling_phonon_en}
\begin{tabular}{cc|ccccc}
\hline
\multicolumn{2}{c|}{Sampling parameters} & \multirow{2}{*}{Relaxed} & \multicolumn{4}{c}{Frequency Threshold (THz)} \\
\cline{1-2} \cline{4-7}
Temperature ($\tau$) & Top-$p$ ($p$) & & $>-0.001$ & $>-0.01$ & $>-0.1$ & $>-1.0$ \\
\hline
0.7 & 0.7 & 1221 & 0.229 & 0.229 & 0.257 & 0.454 \\
0.7 & 1.0 & 1151 & 0.203 & 0.218 & 0.314 & 0.595 \\
1.0 & 0.7 & 1170 & \textbf{0.239} & \textbf{0.243} & 0.318 & 0.545 \\
1.0 & 1.0 & 1089 & 0.193 & 0.209 & \textbf{0.420} & \textbf{0.734} \\
\hline
\end{tabular}
\label{tab:LLM70B}
\end{table}

\section{Summary of Large-Scale Crystal Generation and Dynamical Stability}
\label{app:summary}
\begin{figure}[H]
		\centering  
		\includegraphics[width=0.7\linewidth]{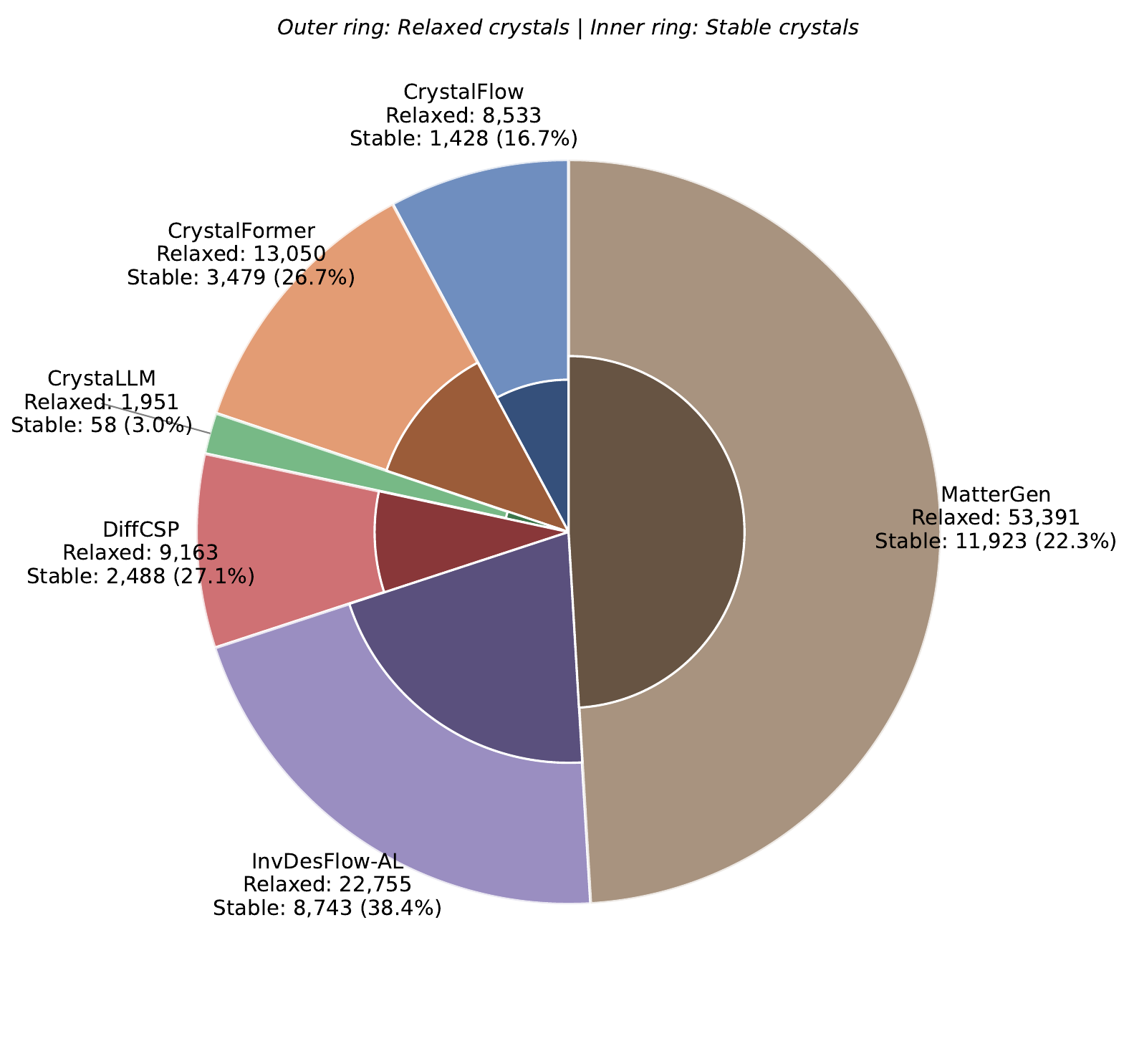}
		\caption{Area-proportional sector plot comparing six crystal generation models in terms of successfully relaxed structures and dynamically stable structures. The sector angle of each model is proportional to the number of relaxed structures, while the area of the darker inner sector represents the fraction of dynamically stable structures identified by phonon calculations. Statistics are aggregated across different training datasets and generation conditions, leading to slight differences from stability rates reported in single-setting main-text analyses.}
		\label{pie-chart} 
\end{figure}
Figure~\ref{pie-chart} presents an area-proportional sector plot summarizing the performance of six crystal generation models in large-scale generation tasks, including the number of successfully relaxed structures and the subset that are dynamically stable. Each sector corresponds to one model, with the sector angle determined by the number of successfully relaxed structures, such that the angular span is proportional to the model’s contribution to the total generated samples, thereby reflecting relative generation scale across models. Within each sector, a darker inner sector with the same angular span but a radius scaled proportionally is overlaid to represent the number of dynamically stable structures, ensuring that the area of the inner sector quantitatively corresponds to the fraction of stable crystals produced by each model. It should be noted that this figure aggregates all structures generated by the same method across different training datasets and generation conditions, and applies a unified phonon-based dynamical stability evaluation to all successfully relaxed structures. Consequently, the dynamical stability rates shown here differ slightly from those reported in the main-text bar plots, which are based on single settings or specific generation conditions. Taking MatterGen as an example, a total of 53,391 newly generated structures were successfully relaxed across all settings, among which 11,923 were identified as dynamically stable, corresponding to a stability rate of 22.3\%.
\section{Convergence Analysis of Dynamical Stability Rate}
\label{app:convergence}
\begin{figure}[H]
		\centering  
		\includegraphics[width=1.0\linewidth]{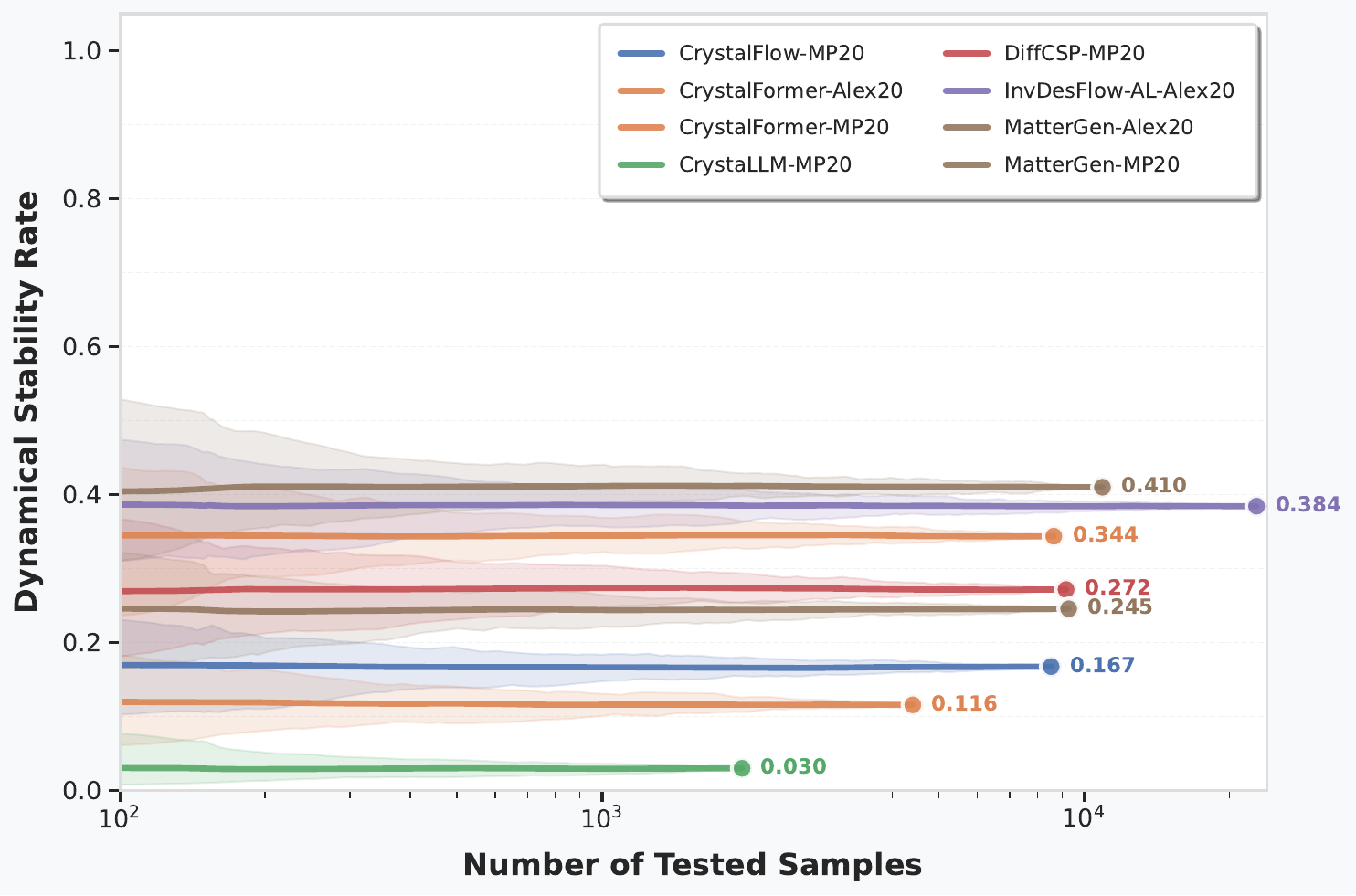}
		\caption{Convergence of the dynamical stability rate for various crystal generative models as a function of the number of tested samples. Shaded regions represent 95\% confidence intervals obtained via 80 iterations of Bootstrap resampling. Stability sequences were simulated by assuming uniform distribution of stable crystals among all tested samples. The figure demonstrates that stability estimates converge as sample size increases, with errors becoming negligible above ~4,000 samples, ensuring robust and fair model ranking.
}
		\label{stable-conv} 
\end{figure}

Figure~\ref{stable-conv} illustrates the convergence of the dynamical stability rate for various crystal generative models as a function of the number of tested samples. To simulate the variation of stability rate with sample size, we generated binary sequences for each model (stable = 1, unstable = 0) based on the total number of tested crystals and the number of successfully relaxed, dynamically stable crystals, assuming that stable crystals are uniformly distributed among the samples. Bootstrap resampling (80 iterations) was then applied to randomly shuffle the sequences and compute the cumulative stability rate, yielding the mean curve and the 95\% confidence interval at each sample size. To smooth the curves, the cumulative ratios were processed using a uniform moving average. The figure clearly shows that as the number of tested samples increases, the stability rate curves for all models gradually converge, and the shaded confidence intervals narrow. This indicates that when the sample size exceeds approximately 4,000, the estimation error of the stability rate becomes sufficiently small, and further increasing the number of samples has negligible impact on the ranking of the models. Although the effective test sample size for CrystaLLM is slightly below this threshold, the error range is still insufficient to alter its final ranking. This analysis provides quantitative support for the comparison of dynamical stability in the main text, ensuring the reliability and fairness of the model performance evaluation.

\section{\textcolor{black}{Summary of Crystal Generation}}
\label{app:generation}
Table~\ref{tab:crystal_generation} summarizes the statistics of crystal generation and stability for the nine crystal generative models evaluated in this study. For each model, the table lists the number of structures that successfully converged during relaxation, the number of dynamically stable crystals identified through phonon calculations, the number of crystals successfully generated via the input scripts, the count of unique CIF files after removing duplicates, and the total number of crystals originally generated. This comprehensive overview provides a quantitative comparison of the performance of different models in terms of generation success, structural relaxation, and dynamical stability, offering a valuable reference for further analysis and discussion in the main text.

\begin{table}[htbp]
\centering
\small
\caption{Summary of crystal generation and stability statistics for different models.}
\label{tab:crystal_generation}
\begin{tabular}{lccccc}
\hline
Model & Relaxed & Dynamically Stable & Input Script Success & Unique CIFs & Total Generated \\
\hline
CrystalFlow-MP20       & 8,533  & 1,428  & 8,852  & 9,952   & 16,000 \\
CrystalFormer-Alex20   & 8,642  & 2,969  & 8,807  & 8,986   & 40,000 \\
CrystalFormer-MP20     & 4,408  & 510    & 4,990  & 5,143   & 20,000 \\
CrystaLLM-small         & 1,951  & 58     & 2,074  & 2,074   & 16,000 \\
\textcolor{black}{CrystaLLM-large-fomula}         & 10,387  & 1913    & 10,387  & 10,902   & 10,902 \\
\textcolor{black}{CrystaLLM-large-sg}         & 9,975  & 1958   & 9,975  & 10,902   & 10,902 \\
\textcolor{black}{LLaMA2-70B}         & 4,633  & 1005   & 4,633  & 4,633   & 4,633 \\
DiffCSP-MP20           & 9,163  & 2,488  & 9,959  & 10,000  & 16,000 \\
InvDesFlow-AL-MP20           & 8,000  & 2176  & -  & -  & - \\
InvDesFlow-AL-Alex20   & 22,755 & 8,743  & 24,997 & 25,000  & 30,000 \\
MatterGen-Alex20       & 10,902 & 4,469  & 11,829 & 11,829  & 16,000 \\
MatterGen-MP20         & 9,279  & 2,278  & 10,000 & 10,000  & 16,000 \\
\hline
\end{tabular}
\end{table}
In Table~\ref{tab:crystal_generation}, “Total Generated” denotes the total number of crystals produced by each model. To ensure feasibility for subsequent relaxation and phonon calculations, the generated crystals were first filtered to remove duplicates and invalid CIF files. If the resulting set exceeded 10,000 crystals, a random subset of 10,000 was selected for relaxation; otherwise, all available crystals were used for phonon calculations. Notably, InvDesFlow-AL-Alex20 and MatterGen-Alex20 represent special test cases with substantially larger generation counts, designed to investigate the sample size required for the convergence of dynamical stability estimates. Experiments showed that stability ratios converge reliably with approximately 4,000 samples, making further increases unnecessary for subsequent analyses.
Since InvDesFlow-AL-MP20 and DiffCSP-MP20 share the same model architecture and MP20 training set, no systematic difference in data distribution is expected. The dynamical stability ratio of InvDesFlow-AL-MP20 reported in the main text is therefore estimated by randomly sampling 8,000 structures from the 9,163 DiffCSP-MP20 samples with completed dynamical stability evaluations and averaging over five independent trials, which matches the sample size while reducing statistical fluctuations and ensuring a fair and robust comparison between models.

\section{Crystal Relaxation and Dynamical Stability Statistics Across Bandgap Ranges}
\label{app:bandgap}
Table~\ref{tab:bandgap_stability} summarizes the statistics of crystal relaxation and dynamical stability across different bandgap ranges. For each bandgap category, the table lists the number of structures that successfully converged during relaxation, the number of dynamically stable crystals identified via phonon calculations, and the total number of crystals generated. This overview provides a quantitative reference for analyzing the relationship between bandgap and dynamical stability.
\begin{table}[htbp]
\centering
\caption{Statistics of crystal relaxation and dynamical stability for different bandgap ranges.}
\label{tab:bandgap_stability}
\begin{tabular}{lccccc}
\hline
Bandgap Range & Relaxed & Dynamically Stable & \textcolor{black}{Input Script Success} & Total Generated \\
\hline
$E_g = 0.5\ \text{eV}$ & 6,524 & 1,532 & \textcolor{black}{7,222} & 10,000 \\
$E_g = 1.5\ \text{eV}$ & 9,478 & 1,448 & \textcolor{black}{10,594} & 16,000 \\
$E_g = 2.5\ \text{eV}$ & 6,133 & 816  & \textcolor{black}{6,916} & 10,000 \\
$E_g = 3.5\ \text{eV}$ & 5,735 & 763  & \textcolor{black}{6,644} & 10,000 \\
$E_g = 4.5\ \text{eV}$ & 5,340 & 617  & \textcolor{black}{6,109} & 10,000 \\
\hline
\end{tabular}
\end{table}

\begin{figure}[H]
		\centering  
		\includegraphics[width=0.5\linewidth]{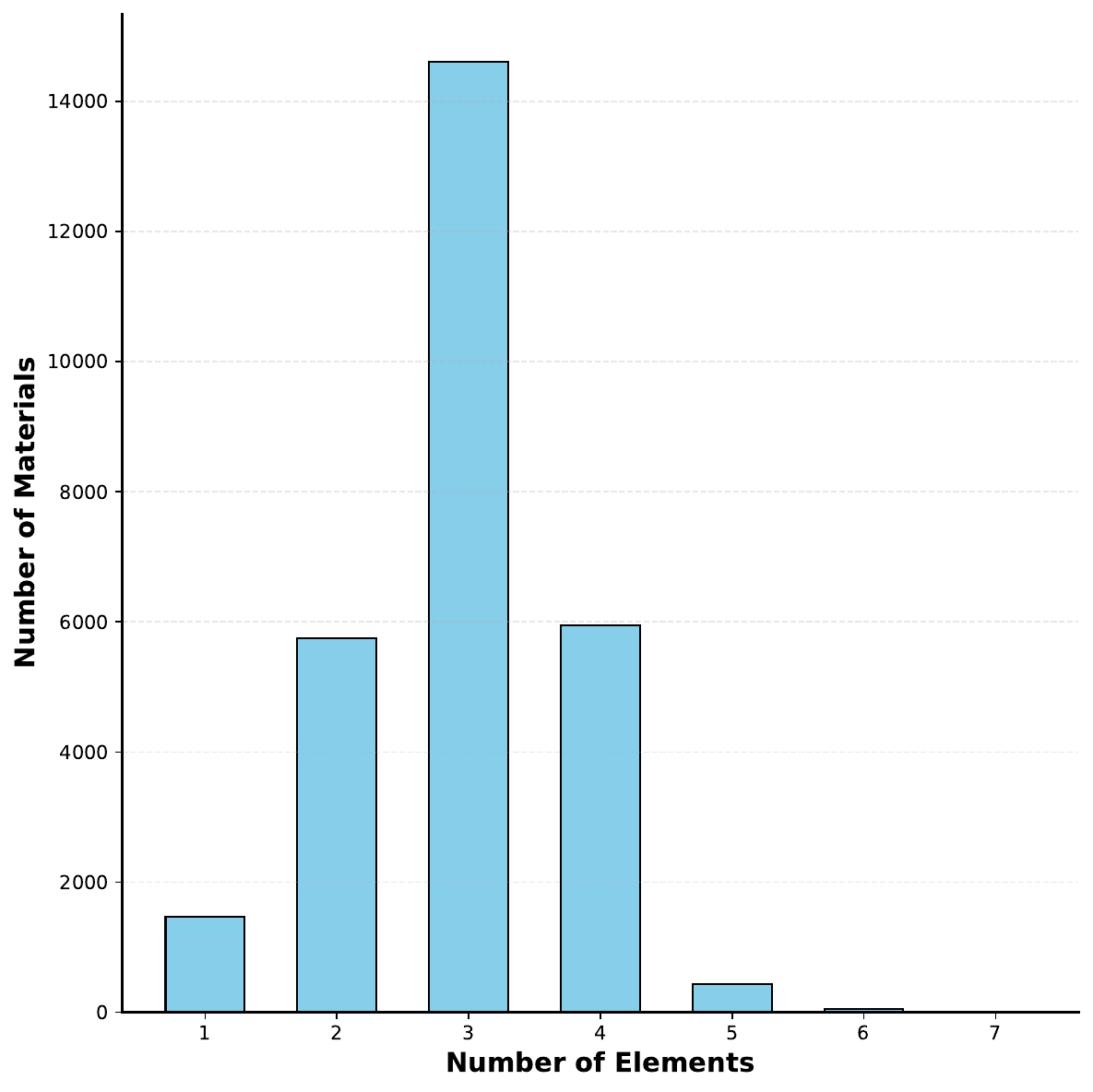}
		\caption{Elemental distribution of 28,119 dynamically stable crystals}
		\label{stable-Elemental-distribution} 
\end{figure}

\section{\textcolor{black}{Data Download}}
\label{app:data}
\textcolor{black}{We have added explicit references and access links for all datasets listed in Table~\ref{tab:datasets}.}

\begin{table}[H]
\small
\color{black}
\centering
\caption{Summary of datasets used in this work. All datasets are publicly available and can be accessed via the provided links.}
\begin{tabular}{lll}
\hline
\textbf{Dataset} & \textbf{Description} & \textbf{Access} \\
\hline

PhononBench (YAML-1) 
& Force constants

& \url{https://zenodo.org/records/19328178} \\

PhononBench (YAML-2) 
& Force constants 

& \url{https://zenodo.org/records/19338671} \\

PhononBench (NPZ) 
& Phonon band structures

& \url{https://zenodo.org/records/19317118} \\

Relaxed Crystals
& Relaxed crystal structures 

& \url{https://zenodo.org/records/18185662} \\

Alex20 
& Training datasets

& \url{https://zenodo.org/records/19346093} \\

MP20 
& Training datasets 
& \url{https://zenodo.org/records/19346093} \\

DFT-120Materials
& DFT\&uMLIPs
& \url{https://zenodo.org/records/19395921} \\

\hline
\end{tabular}
\label{tab:datasets}
\end{table}



\end{document}